\documentclass[prd,preprint,tightenlines,floatfix,
showpacs,preprintnumbers,nofootinbib,eqsecnum,superscriptaddress]{revtex4}

\usepackage{amsmath,amsfonts,amssymb,amstext,mathrsfs,amscd}
\usepackage{graphicx}
\usepackage{mathpazo}
\usepackage{subfigure}
\usepackage{epsf,float}
\usepackage{revsymb}
\usepackage{bm}
\usepackage{dcolumn}
\usepackage{braket}
\usepackage{color,xcolor}
\usepackage{soul}
\usepackage{graphicx}
\usepackage{subfigure}
\usepackage{multirow}
\usepackage{tabularx}
\usepackage{pstricks}
\usepackage[section]{placeins}
\usepackage{booktabs}
\usepackage{array}
\usepackage{hyperref}

\begin{document}

\begin{flushright}
LU TP 16-XX\\
March 2016
\vskip1cm
\end{flushright}

\title{Can the diphoton enhancement at 750 GeV be due to a neutral technipion?}

\author{Piotr Lebiedowicz}
\email{Piotr.Lebiedowicz@ifj.edu.pl}
\affiliation{Institute of Nuclear Physics, Polish Academy of Sciences, PL-31-342 Krak\'ow, Poland}

\author{Marta {\L}uszczak}
\email{luszczak@univ.rzeszow.pl}
\affiliation{Department of Theoretical Physics, University of Rzesz{\'o}w, PL-35-959 Rzesz{\'o}w, Poland}

\author{Roman Pasechnik}
\email{Roman.Pasechnik@thep.lu.se}
\affiliation{Department of Astronomy and Theoretical Physics, Lund University, 
SE-223 62 Lund, Sweden}

\author{Antoni Szczurek
\footnote{Also at University of Rzesz{\'o}w, PL-35-959 Rzesz{\'o}w, Poland.}}
\email{Antoni.Szczurek@ifj.edu.pl}
\affiliation{Institute of Nuclear Physics, Polish Academy of Sciences, PL-31-342 Krak\'ow, Poland}

\begin{abstract}
We discuss a scenario in which the diphoton enhancement at $M_{\gamma \gamma}$ = 750 GeV, 
observed by the ATLAS and CMS Collaborations, is a neutral technipion $\tilde{\pi}^0$. 
We consider two distinct minimal models
for the dynamical electroweak symmetry breaking. 
In a first one, two-flavor vector-like technicolor (VTC) model, 
we assume that the two-photon fusion is a dominant production mechanism. 
We include $\gamma \gamma \to {\tilde \pi}^0$ and
production of technipion associated with one or two jets.
All the considered mechanisms give similar contributions.
With the strong Yukawa (technipion-techniquark) coupling
$g_{TC}$ = 10 - 20 we obtain the measured cross section of the ``signal''. 
With such values of $g_{TC}$ we get a relatively small $\Gamma_{\rm tot}$. 
In a second approach, one-family walking technicolor (WTC) model,
the isoscalar technipion is produced dominantly via the gluon-gluon fusion. 
We also discuss the size of the signal at lower energies (LHC, Tevatron) 
for $\gamma \gamma$ (VTC) and jet-jet (WTC) final states
and check consistency with the existing experimental data. 
We predict a measurable cross section for ${\tilde \pi}^0$ production 
associated with one or two soft jets. 
The technipion signal in both models 
is compared with the SM background diphoton contributions. 
We observe the dominance of inelastic-inelastic processes 
for $\gamma \gamma$ induced processes. 
In the VTC scenario, we predict the signal cross section for purely exclusive $p p \to p p \gamma
\gamma$ processes at $\sqrt{s}$ = 13 TeV to be about 0.2 fb. Such a cross section 
would be, however, difficult to measure with the planned integrated luminosity. 
In all considered cases the signal is below the
background or/and below the threshold set by statistics.
\end{abstract}


\maketitle

\section{Introduction}
\label{sec:Introduction}

Recently both the ATLAS and CMS Collaborations announced an observation of an intriguing enhancement in the diphoton invariant mass at 
$M_{\gamma \gamma}\approx750$~GeV in proton-proton collisions at $\sqrt{s} =13$~TeV \cite{ATLAS:2015,CMS:2015dxe}
\footnote{For a search 
of diphoton resonances at smaller mass range $65 < M_{\gamma \gamma}< 600$~GeV and $\sqrt{s} =8$~TeV, see Ref.~\cite{Aad:2014ioa}.}. 
Remarkably, such a hint to a possible New Physics signal has triggered a lot of research activities in recent months looking for its possible interpretation 
in various theoretical scenarios of New Physics (see e.g. Refs~\cite{Ellis:2015oso,Franceschini:2015kwy,Csaki:2015vek,Fichet:2015vvy,Fichet:2016pvq}). 
However, before one could be certain about a possible nature of such a hint, it requires further confirmation by collecting a better statistics. 
If it is confirmed it will be a very important discovery related to first observation of the signal beyond Standard Model (SM). Different scenarios 
are possible a priori. The resonance signal observed means that the potentially new state decays into (and thus should couple to) two photons.
What is the dominant production mechanism is a speculation at this stage. 
Several options are possible a priori. In one of them gluon-gluon fusion
is the dominant production mechanism. If coupling of the new state to
gluons is weak other options have to be considered. 
In the present analysis we consider such an example.
The two-photon induced production of various objects became recently 
rather topical. 
This includes, for instance, production of $c \bar{c}$ \cite{Luszczak:2011uh}, $b \bar{b}$ \cite{Maciula:2010vc,Maciula:2010tv}, $l^+ l^-$ 
\cite{daSilveira:2014jla,Luszczak:2015aoa}, $W^+ W^-$ \cite{Lebiedowicz:2012gg,Luszczak:2014mta} 
or $H^+ H^-$ \cite{Lebiedowicz:2015cea}. In the current study,
we continue this line of research and consider the two-photon production mechanism of a new lightest composite state, the technipion predicted
by various technicolor models, thus probing its potential to explain the 750 GeV excess.

A high-scale strongly-coupled physics can, in principle, be responsible for the dynamical electro-weak symmetry breaking (EWSB) 
in the SM by means of strongly-interacting technifermion condensation (see e.g. Refs~\cite{Weinberg:1975gm,Susskind:1978ms,Eichten:1979ah}). 
Such a dynamics typically predicts a plenty of new composite states close to the EWSB scale, in particular, relatively light composite scalar Higgs-like 
particles and pseudoscalar technipions. A consistent realisation of the underlined compositeness scenarios is typically limited by the precision 
SM tests \cite{Peskin:1990zt,Peskin:1991sw,Galloway:2010bp} and the ongoing SM-like Higgs boson studies \cite{Arbey:2015exa} 
(for a detailed review see e.g. Refs~\cite{Hill:2002ap,Sannino:2009za}). One of the appealing and consistent classes of TC models with 
a vector-like (Dirac) UV completion is known as the vector-like TC (VTC) scenario~\cite{Kilic:2009mi}.
The simplest version of the VTC scenario applied to the EWSB possessed two Dirac techniflavors and 
a SM-like Higgs boson~\cite{Pasechnik:2013bxa,Lebiedowicz:2013fta,Pasechnik:2014ida}. Recently, the concept of 
Dirac UV completion has also emerged in composite Higgs boson scenarios with confined $SU(2)_{\rm TC}$ symmetry 
\cite{Cacciapaglia:2014uja,Hietanen:2014xca}. 

\begin{figure}[!ht]
\includegraphics[width=0.2\textwidth]{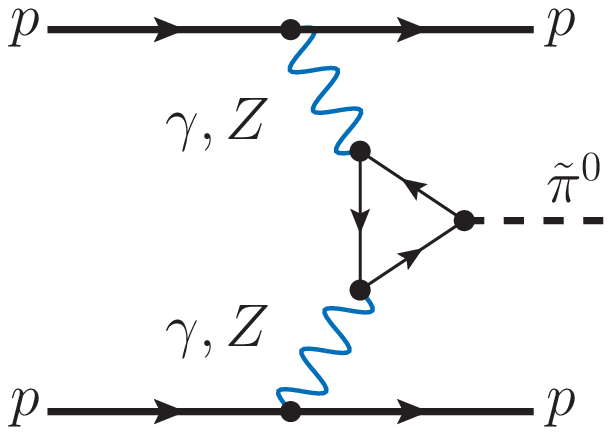}
\includegraphics[width=0.2\textwidth]{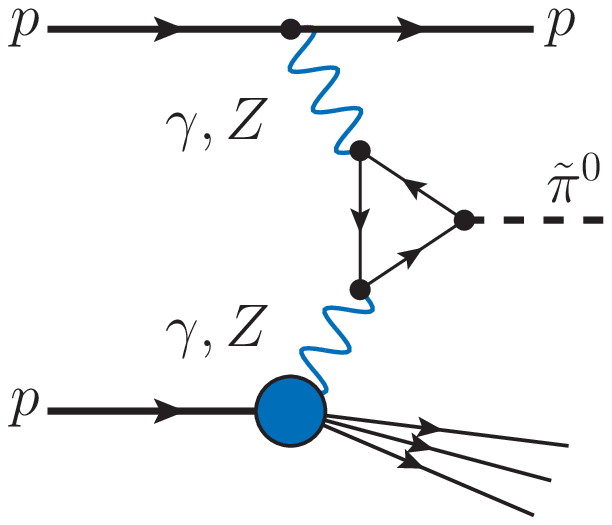}
\includegraphics[width=0.2\textwidth]{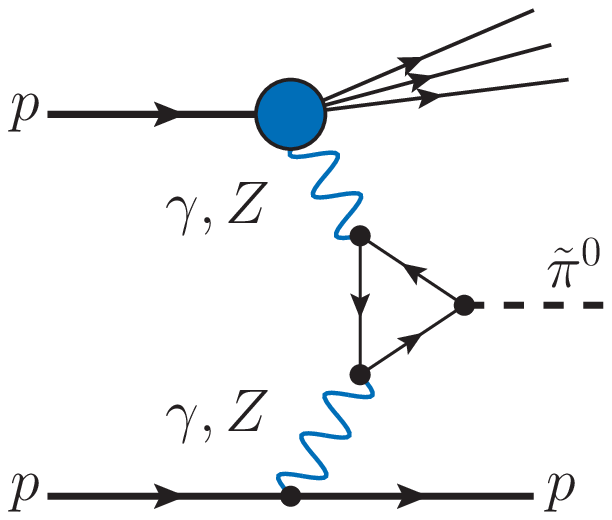}
\includegraphics[width=0.2\textwidth]{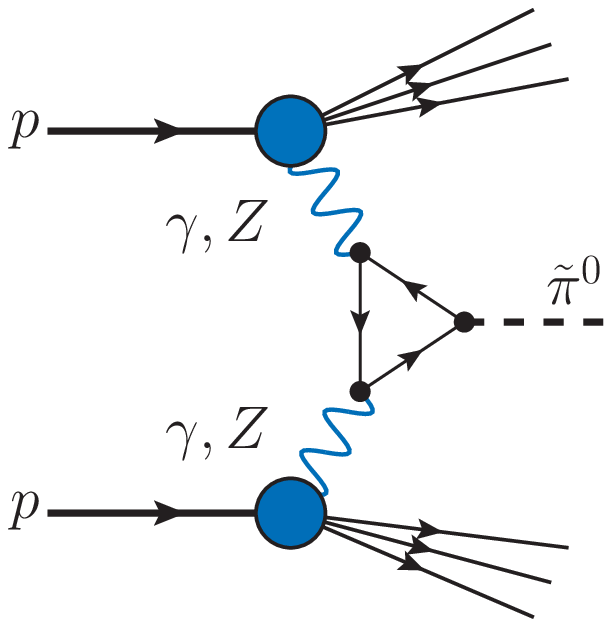}
  \caption{\label{fig:diagrams_order1}
  \small
Diagrams of neutral technipion production 
via the $\gamma \gamma$, $\gamma Z$ and $ZZ$ fusion 
in $pp$-collisions.}
\end{figure}

\begin{figure}[!ht]
\includegraphics[width=0.24\textwidth]{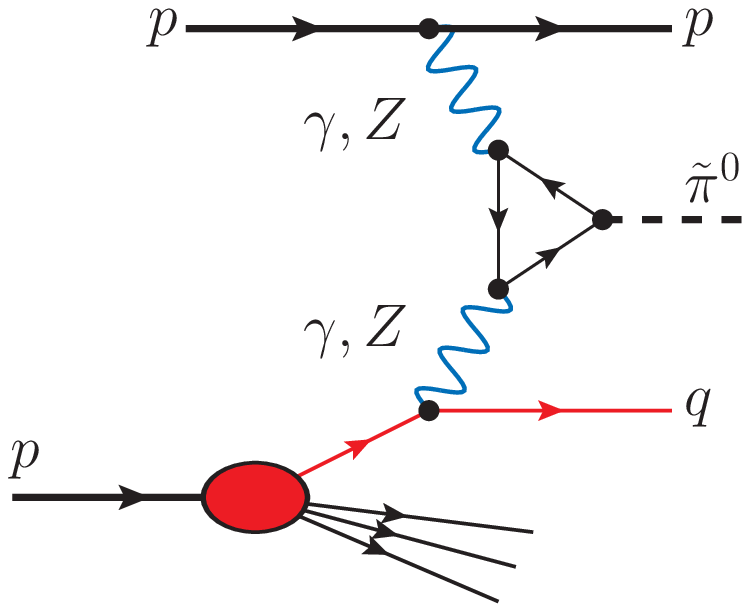}
\includegraphics[width=0.24\textwidth]{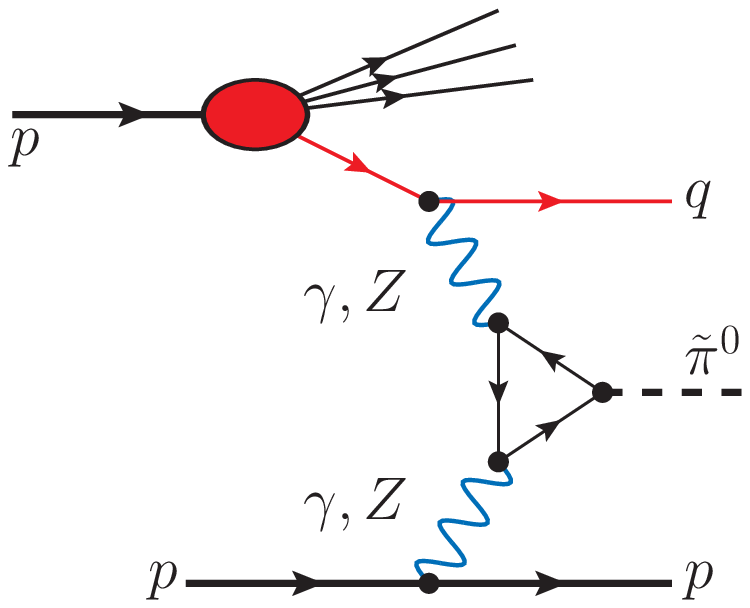}
\includegraphics[width=0.24\textwidth]{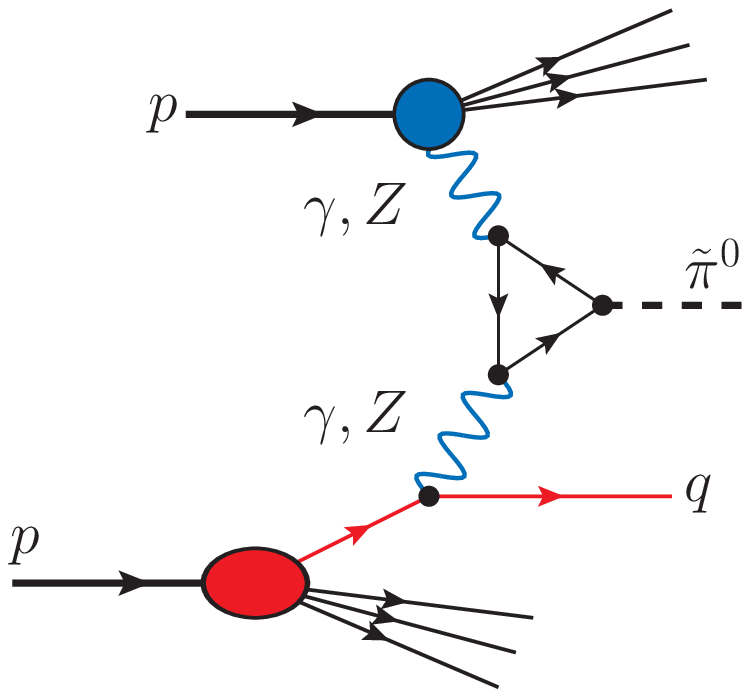}
\includegraphics[width=0.24\textwidth]{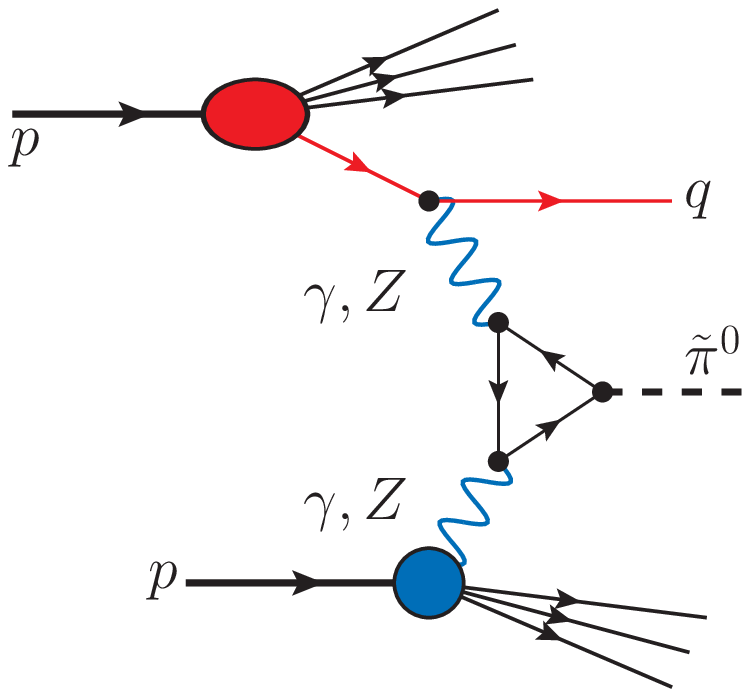}
  \caption{\label{fig:diagrams_order2}
  \small
Technipion production via the $2 \to 2$ partonic subprocesses. }
\end{figure}

Below, we discuss possible implications of the neutral pseudoscalar technipion 
in the two-flavor VTC \cite{Pasechnik:2013bxa} and one-family 
Walking TC (WTC) \cite{Matsuzaki:2015che} scenarios of the dynamical EWSB for the diphoton 750 GeV signature at the LHC. In the simplest VTC scenario, 
the technipion does not couple to quarks and gluons and is thus produced only via EW vector ($\gamma$ and $Z$) 
boson fusion (VBF) mechanism with a slight dominance of the $\gamma\gamma$ fusion channel at large invariant masses. 
The corresponding technipion production processes can be classified into three groups depending on the QED order. The first group 
of diagrams is shown in Fig.~\ref{fig:diagrams_order1}. Diagrams in Fig.~\ref{fig:diagrams_order2} show the higher QED-order group,
that is, the technipion production associated with one jet. We consider here the $\gamma q \to {\tilde \pi}^0 q$ and $q \gamma \to {\tilde \pi}^0 q$ 
subprocesses. In even higher QED-order we have to include also $q_{i} q'_{j} \to q_{i} {\tilde \pi}^0 q'_{j}$ subprocesses, where $q_{i}$ and $q_{j}$ 
can be either a quark or an antiquark of various flavours from each of the colliding protons. In the Walking TC scenario, the technipion is produced 
dominantly by the ordinary gluon-gluon fusion (relevant also for the Higgs boson production at the LHC) and can decay with sizeable branching fraction 
into two-photon final state\footnote{Other distinct scenarios were explored recently in Refs~\cite{Molinaro:2015cwg,Pilaftsis:2015ycr,Mambrini:2015wyu} 
and will not be discussed here.}.
\begin{figure}[!ht]
\includegraphics[width=0.25\textwidth]{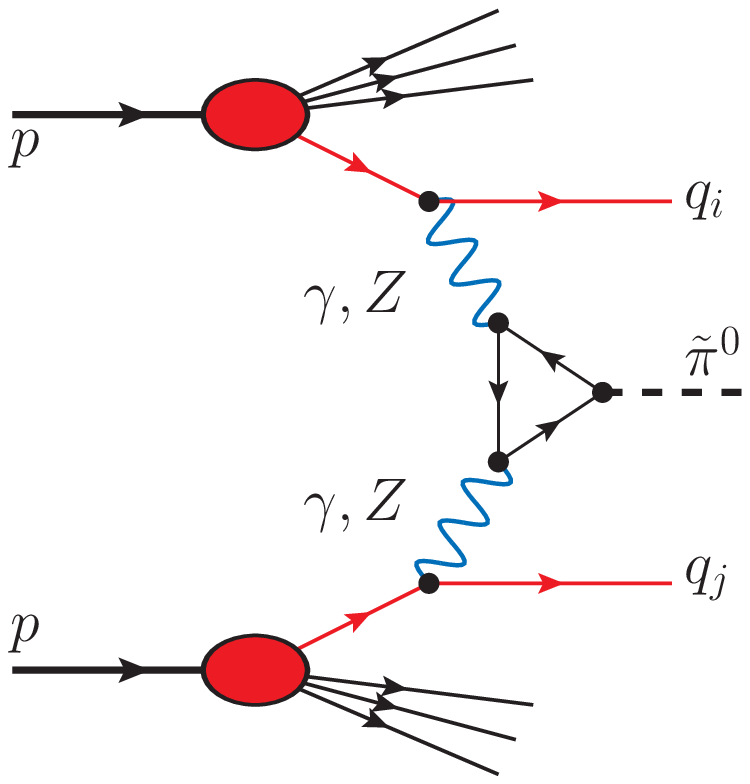}
  \caption{\label{fig:diagrams_order3}
  \small
Technipion production via the $2 \to 3$ partonic subprocesses. }
\end{figure}

\section{Vector-like TC model: a short overview}

Before we go to the production mechanisms relevant for the neutral technipion production, 
let us summarize the main points of the VTC model.

The Dirac techniquarks in the minimal two-flavor VTC model form pseudoreal representations of the global $SU(4)$ group which 
contains the chiral $SU(2)_L\otimes SU(2)_R$ symmetry in the technimeson sector. In the linear realisation of the VTC model, 
at low energies the global chiral $SU(2)_L\otimes SU(2)_R$ symmetry describes the effective interactions the lightest technimeson 
states, technipion $P^a$ ($a=1,2,3$), technisigma $S$ and constituent Dirac techniquarks ${\tilde Q}_\alpha$ $\alpha=1,2$ 
\cite{Pasechnik:2013bxa,Pasechnik:2014ida}, similar to those in quark-meson effective theories of QCD hadron physics \cite{Tetradis:2003qa}. 

For the Dirac UV completion, the vector subgroup of the global chiral group $SU(2)_{V\equiv L+R}$ and the local weak-isospin 
symmetry of the SM $SU(2)_W$ are locally isomorphic. Therefore, the representations of the $SU(2)_{V}$ can always be mapped 
onto the representations of $SU(2)_W$ such that the $P^a$, $S$ and Dirac ${\tilde Q}_\alpha$ can be classified as triplet,
singlet and doublet representations of the gauge $SU(2)_W$ symmetry. Such a straightforward way of introducing weak 
interactions into the technimeson sector was proposed in the framework of the VTC model in Ref.~\cite{Pasechnik:2013bxa}.
In particular, it was demonstrated for the first time that practically any simple Dirac UV completion with chirally-symmetric 
weak interactions naturally escapes the electroweak precision constraints which is considered the basic motivation 
for the VTC scenario despite its simplicity.

Let us consider a single $SU(2)_{\rm W}$ doublet of Dirac techniquarks 
 \begin{eqnarray} 
 \label{Tdoub}
 {\tilde Q} = \left(
      \begin{array}{c}
         U \\
         D
      \end{array}
             \right)\,, 
 \end{eqnarray}
confined under $SU(N_{\rm TC})_{\rm TC}$ at an energy scale $\Lambda_{\rm TC}$ above the EW scale. For a QCD-like scenario
we choose $N_{\rm TC}=3$ and the hypercharge $Y_{\tilde Q}=1/3$ provided that electric charges of corresponding bounds states are 
integer-valued. Then, the phenomenological interactions of the constituent techniquarks and the lightest technimesons are described by 
the (global) chiral $SU(2)_{\rm R}\otimes SU(2)_{\rm L}$ invariant low-energy effective Lagrangian in the linear $\sigma$-model (L$\sigma$M)
 \begin{eqnarray} 
 \nonumber
  && {\cal L}_{{\rm L} \sigma {\rm M}} = \frac12\, \partial_{\mu} S\,
  \partial^{\mu} S +
  \frac12 D_{\mu} P_a\, D^{\mu} P_a + i \bar{{\tilde Q}}\hat{D}{\tilde Q} \\
  && -\, g_{\rm TC} \bar{{\tilde Q}}(S+i\gamma_5\tau_a P_a){\tilde Q}
  - g_{\rm{TC}}\,S\,\langle\bar{{\tilde Q}}{\tilde Q}\rangle  \nonumber \\
  && -\,\lambda_{\rm H}{\mathcal{H}}^4
  - \frac14\lambda_{\rm TC}(S^2+P^2)^2 +
  \lambda{\mathcal{H}}^2(S^2+P^2) \nonumber \\
  && +\, \frac12\mu^2_{\rm S}(S^2+P^2)+\mu_{\rm H}^2{\mathcal{H}}^2 \,,
  \label{LsM}
 \end{eqnarray}
where the ``source'' term linear in technisigma is proportional to the flavor-diagonal techniquark condensate 
$\langle\bar{{\tilde Q}}{\tilde Q}\rangle<0$, ${\mathcal{H}}^2=\mathcal{H}\mathcal{H}^\dagger$,
$P^2\equiv P_aP_a={\tilde \pi}^{0}{\tilde \pi}^{0}+2{\tilde \pi}^+{\tilde \pi}^-$, and the
corresponding covariant derivatives read
 \begin{eqnarray} 
 \nonumber
    &&\hat{D}{\tilde Q} = \gamma^{\mu} \left( \partial_{\mu}
       - \frac{iY_{{\tilde Q}}}{2}\, g'B_{\mu} - \frac{i}{2}\,
       g W_{\mu}^a \tau_a \right)\tilde{Q}\,, \\
    &&D_{\mu} P_a = \partial_{\mu} P_a + g
\epsilon_{abc} W^b_{\mu} P_c\,. \label{DQ}
 \end{eqnarray}
For simplicity, the Higgs boson doublet $\mathcal{H}$ is kept to be elementary.
The choice of the ``source'' term in Eq.~(\ref{LsM}) is rather natural since it (a) induces a pseudo-Goldstone 
mass scale for pseudo-Goldstone technipion $m_{\tilde \pi}$, and (b) relates the scales of the spontaneous EW 
and chiral symmetry breakings as well as the constituent techniquark mass scale $m_{\tilde Q}$ with the value of 
the techniquark condensate \cite{Tetradis:2003qa,Pasechnik:2013bxa}.

In the conformal limit of the theory $\mu_{S,H}\ll m_{\tilde \pi}$, 
the EW and chiral symmetries are broken 
by the Higgs $v\simeq 246\,{\rm GeV}$ and technisigma $u$ vevs 
\begin{eqnarray} \nonumber
 && \mathcal{H} = \frac{1}{\sqrt{2}}\left(
\begin{array}{c}
\sqrt{2}i\phi^-  \\
H+i\phi^0
\end{array}\right)\,,\quad \langle H\rangle \equiv v \,, 
\quad \langle S\rangle \equiv u \gtrsim v \,, \\
 && H=v+h c_\theta-\tilde \sigma s_\theta\,, \qquad  S = u + h s_\theta+\tilde \sigma c_\theta\,,
\label{shifts}
\end{eqnarray}
respectively, which are initiated by the techniquark condensation in the confined regime, i.e.
 \begin{eqnarray} \nonumber
&&u=\left(\frac{g_{\rm{TC}}\lambda_{\rm H}}{\delta}\right)^{1/3}
|\langle\bar{{\tilde Q}}{\tilde Q}\rangle|^{1/3}\,, \\
&&v=\left(\frac{|\lambda|}{\lambda_{\rm H}}\right)^{1/2}
\left(\frac{g_{\rm{TC}}\lambda_{\rm H}}{\delta}\right)^{1/3}
|\langle\bar{{\tilde Q}}{\tilde Q}\rangle|^{1/3}\,. \label{u-v-min}
 \end{eqnarray}
Then, technipions and techniquarks acquire a dynamical effective mass
 \begin{eqnarray*}
&& m_{\pi}^2=-\frac{g_{\rm TC} \langle\bar{{\tilde Q}}{\tilde Q}\rangle}{u}\,, \\
&& m_U=m_D\equiv m_{\tilde Q} = g_{\rm TC}\, u \,.
 \end{eqnarray*}
Above, $s_\theta\equiv \sin\theta$, $c_\theta\equiv \cos\theta$, $\delta=\lambda_{\rm H}\lambda_{\rm{TC}} - \lambda^2$,
$g_{\rm{TC}}>0$ and $\lambda_H>0$. The phenomenologically consistent regime corresponds to a small 
Higgs-technisigma mixing $\theta \ll 1$ which is realised in the TC decoupling limit $v/u\ll 1$ \cite{Pasechnik:2013bxa}. 
As a characteristic feature of the model, the technipions in the VTC model can stay relatively light and do not 
have tree-level couplings to the SM fermions, so can only be produced in vector-boson fusion channels 
\cite{Pasechnik:2013bxa,Lebiedowicz:2013fta}. In what follows, we discuss possible signatures of 
the VTC technipions at the LHC.

\vspace{0.5cm}
\textbf{Decay widths for the neutral technipion in the vector-like TC model}
\vspace{0.2cm}

The techniquark-loop amplitude has the following form \cite{Pasechnik:2013bxa}
\begin{eqnarray}
 && i{\cal V}_{\tilde{\pi}^0\,V_1\,V_2} =
 F_{V_1 V_2}(m_1^2,m_2^2,m^2_{\tilde{\pi}^0};m_{\tilde{Q}}^2)
 \epsilon_{\mu \nu \rho \sigma} p_1^{\mu} p_2^{\nu}
 {\varepsilon^*_1}^{\rho} {\varepsilon^*_2}^{\sigma}\,,\\
 && F_{V_1 V_2}=\frac{N_{\rm TC}}{2\pi^2}\sum_{\tilde{Q}=U,D} g_{V_1}^{\tilde{Q}}\,
 g_{V_2}^{\tilde{Q}}\,g_{\tilde{\pi}^0}^{\tilde{Q}}\,m_{\tilde{Q}}\,
 C_0(m_1^2,m_2^2,m_{\tilde \pi}^2;m_{\tilde{Q}}^2)\,,
\label{FV1V2}
\end{eqnarray}
where $C_0(m_1^2,m_2^2,m_3^2;m^2)\equiv
C_0(m_1^2,m_2^2,m_3^2;m^2,m^2,m^2)$ is the standard finite
three-point function, $p_{1,2}$, $\varepsilon_{1,2}$ and
$M_{1,2}$ are the four-momenta, polarization vectors of the vector
bosons $V_{1,2}$ and their on-shell masses, respectively, and
neutral technipion couplings to $U,D$ techniquarks are
\begin{eqnarray}
g_{\tilde{\pi}^0}^U=g_{\rm TC}\,,\quad g_{\tilde{\pi}^0}^D=-g_{\rm TC}\,,
\end{eqnarray}
while gauge couplings of techniquarks $g_{V_{1,2}}^{\tilde{Q}}$ are defined in
Ref.~\cite{Pasechnik:2013bxa}. Finally, the explicit expressions of the effective neutral technipion 
couplings $F_{V_1 V_2}$ for on-shell $V_1 V_2=\gamma\gamma$, $\gamma Z$ and $ZZ$
final states are\footnote{We should notice here that the $\tilde{\pi}^0 \to W^{+}W^{-}$ decay mode 
is forbidden by symmetry \cite{Pasechnik:2013bxa}.}
\begin{eqnarray}
&&F_{\gamma\gamma}=
\frac{4\alpha_{em}\,g_{\rm TC}}{\pi}\frac{m_{\tilde{Q}}}{m_{\tilde{\pi}^{0}}^2}\,
\arcsin^2\Bigl(\frac{m_{\tilde{\pi}^{0}}}{2m_{\tilde{Q}}}\Bigr)\,, \qquad
\frac{m_{\tilde{\pi}^{0}}}{2m_{\tilde{Q}}}<1\,,
\label{F_gamgam}\\
&&F_{\gamma Z}=
\frac{4\alpha_{em}\,g_{\rm TC}}{\pi}\frac{m_{\tilde{Q}}}{m_{\tilde{\pi}^{0}}^2}\,\cot2\theta_W\,
\Big[\arcsin^2\Bigl(\frac{m_{\tilde{\pi}^{0}}}
{2m_{\tilde Q}}\Bigr)-\arcsin^2\Bigl(\frac{m_Z}{2m_{\tilde{Q}}}\Bigr)\Big]\,,
\label{F_gamZ}\\
&&F_{ZZ}=
\frac{2\alpha_{em}\,g_{\rm TC}}{\pi}m_{\tilde{Q}}\,
C_0(m_Z^2,m_Z^2,m_{\tilde{\pi}^{0}}^2;m_{\tilde Q}^2)\,,
\label{F_ZZ}
\end{eqnarray}
where $\alpha_{em} = e^{2}/(4\pi)$ is the fine structure constant.

Now the two-body technipion decay width in a vector boson channel
can be represented in terms of the effective couplings (\ref{FV1V2})
as follows:
\begin{eqnarray}
\Gamma(\tilde{\pi}^0\to V_1\,V_2)=r_V\frac{m_{\tilde
\pi}^3}{64\pi}\,\bar{\lambda}^3(m_1,m_2; m_{\tilde
\pi})\,|F_{V_1V_2}|^2\,,
\end{eqnarray}
where $r_V=1$ for identical bosons $V_1$ and $V_2$ and $r_V=2$ for
different ones, and $\bar{\lambda}$ is the normalized K\"all{\'e}n function
\begin{eqnarray}
\bar{\lambda}(m_a,m_b;q)=\Big(1-2\frac{m_a^2+m_b^2}{q^2}+\frac{(m_a^2-m_b^2)^2}{q^4}\Big)^{1/2}\,.
\label{Kallen}
\end{eqnarray}

\begin{figure}[!ht]
\includegraphics[width=0.6\textwidth]{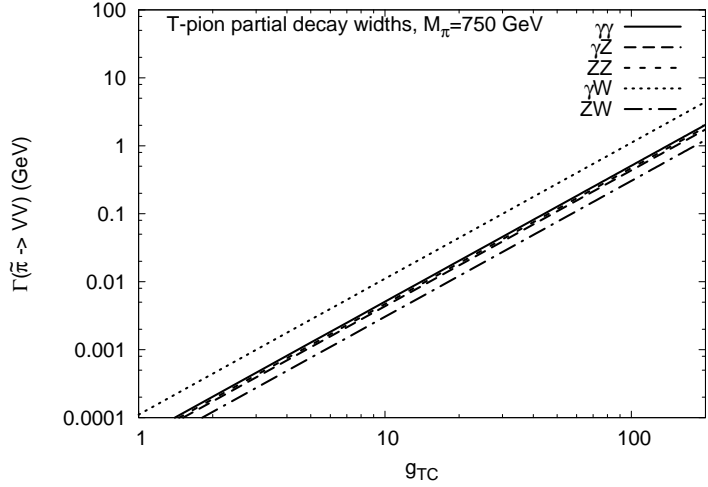}
  \caption{\label{fig:decay_width}
  \small
Decay width in GeV as a function of $g_{TC}$ coupling constant.
The $\gamma W$ and $ZW$ final states are only for charged technipions.
Here we have used our benchmark choice $m_{\tilde Q} = 0.75\, m_{\tilde \pi}^{0}$. }
\end{figure}

For example, in the VTC model, 
for $g_{TC}$ = 10 and $m_{\tilde Q} = 0.75 \,m_{{\tilde \pi}^0}$ one gets:
\begin{eqnarray}
\Gamma(\tilde{\pi}^0 \to \gamma \gamma) &=& 5.136 \times 10^{-3} \;{\rm GeV} \,, \\
\Gamma(\tilde{\pi}^0 \to \gamma Z) &=& 4.376 \times 10^{-3} \;{\rm GeV} \,, \\
\Gamma(\tilde{\pi}^0 \to Z Z) &=& 4.734 \times 10^{-3} \;{\rm GeV} \,. 
\label{decay_widths}
\end{eqnarray}
The total decay width is a sum of the tree contributions:
\begin{equation}
\Gamma_{tot} = \Gamma(\tilde{\pi}^0 \to \gamma \gamma)
             + \Gamma(\tilde{\pi}^0 \to \gamma Z)
             + \Gamma(\tilde{\pi}^0 \to Z Z) \; .
\label{total_decay_width}
\end{equation}
The corresponding branching fractions are:
\begin{eqnarray}
Br(\tilde{\pi}^0 \to \gamma \gamma) &=& 0.36 \,, \\
Br(\tilde{\pi}^0 \to \gamma Z) &=& 0.31 \,, \\
Br(\tilde{\pi}^0 \to Z Z) &=& 0.33 \,.
\label{numerical_branching_fractions}
\end{eqnarray}
How the total decay width depends on the model coupling constant $g_{TC}$ is shown in Fig.~\ref{fig:decay_width}.
Only at very large $g_{TC}$ = 300-400 one can reproduce the ATLAS quasi-experimental value $\Gamma_{tot} \simeq$ 45 GeV 
\cite{ATLAS:2015,CMS:2015dxe} (only the ATLAS collaboration claims that
the observed state is broad). Such a gigantic value is far too big compared to an analogical coupling in effective 
quark-hadron interactions in low-energy QCD such that the model looses its physical sense. However, the extraction of 
the total width from the ``experimental data'' is highly speculative and
biased and thus should not be taken too seriously.

\section{Production mechanisms of neutral technipion}
\label{sec:Production_mechanisms}

As mentioned in the introduction we shall consider contributions of different QED-orders 
to the production of hypothetical technipions as shown in 
Figs.~\ref{fig:diagrams_order1}-\ref{fig:diagrams_order3}.
Let us briefly discuss all the contributions one by one.

As already discussed above in our VTC model \cite{Pasechnik:2013bxa} the technipion couples only 
to photons and $Z$ bosons. In the present paper, we shall include only coupling to photons as far as 
the production mechanism of technipion is considered. The couplings to $Z$ bosons do not affect 
the observables significantly and will be considered elsewhere.

\subsection{$2 \to 1$ subprocess}
\label{sec:2to1_mechanism}

The corresponding diagrams in Fig.~\ref{fig:diagrams_order1} can be categorized into three groups 
(elastic-elastic, elastic-inelastic, inelastic-inelastic) depending 
whether the protons survive intact or undergo electromagnetic dissociation.

The cross section for the $2 \to 1$ contribution via the subprocess $\gamma \gamma \to \tilde{\pi}^{0}$ 
can be easily written in the compact form:
\begin{eqnarray}
\frac{d \sigma_{pp \to {\tilde \pi}^0}}{dy_{\tilde{\pi}^{0}}} 
= \frac{\pi}{m_{\tilde{\pi}^{0}}^4} \sum_{i,j}
x_1 \gamma^{(i)}(x_1,\mu_F^2) x_2 \gamma^{(j)}(x_2,,\mu_F^2) 
\overline{|{\cal M}_{\gamma \gamma \to {\tilde \pi}^0}|^2} \,,
\label{first_order}
\end{eqnarray}
where indices $i$ and $j$ denote $i, j = $ el or in, i.e. they correspond to elastic or inelastic flux
($x$-distribution) of equivalent photons, respectively.
The elastic photon flux can be calculated using e.g. Drees-Zeppenfeld parametrization \cite{Drees:1994zx}. 
The factorization scale makes sense only in the case of dissociation of a proton.
Here we take $\mu_F^2 = m_{\tilde{\pi}^{0}}^2$.
Above $y_{\tilde{\pi}^{0}}$ is rapidity of the technipion and
\begin{eqnarray}
x_1 = \frac{m_{\tilde{\pi}^{0}}}{\sqrt{s}} \exp( y_{\tilde{\pi}^{0}})\,, \quad  
x_2 = \frac{m_{\tilde{\pi}^{0}}}{\sqrt{s}} \exp(-y_{\tilde{\pi}^{0}})\,.
\end{eqnarray}
In the leading-order collinear approximation the technipion is produced 
with zero transverse momentum.
To calculate inelastic contributions we use collinear approach with photon PDFs \cite{Martin:2004dh}. 

The matrix element squared has been calculated with effective $\gamma \gamma \to \tilde{\pi}^0$ vertex.
In general, the form factor (\ref{F_gamgam}) describes the coupling of two photons to the technipion resonance
could depend on virtualities of photons. 

\subsection{$2 \to 2$ subprocess}
\label{sec:2to2_mechanism}

Now we discuss diagrams shown in Fig.~\ref{fig:diagrams_order2} with the $\gamma q \to \tilde{\pi}^{0} q$ 
and $q \gamma \to \tilde{\pi}^{0} q$ subprocesses. The cross section can
be also written in a compact way which allows easy calculation of differential distributions:
\begin{eqnarray}
\frac{d \sigma}{d y_{3} d y_{4} d^2 p_{t,\tilde{\pi}^{0}}}
&=& \frac{1}{16 \pi^2 {\hat s}^2}
\sum_{i} x_1 \gamma^{(i)}(x_1,\mu_F^2) x_2 q_{eff}(x_2,\mu_F^2)
\overline{|
{\cal M}_{\gamma q \to \tilde{\pi}^{0} q} |^2} \; , \nonumber \\
&+& \frac{1}{16 \pi^2 {\hat s}^2}
\sum_{j} x_1 q_{eff}(x_1,\mu_F^2) x_2 \gamma^{(j)}(x_2,\mu_F^2)
\overline{|
{\cal M}_{q \gamma \to \tilde{\pi}^{0} q} |^2} \,,
\label{second_order}
\end{eqnarray}
where index ``3'' refers to technipion and index ``4'' refers to outgoing quark/antiquark
and
\begin{eqnarray}
&&x_1 = \frac{m_{1 \perp}}{\sqrt{s}} \exp( y_{3}) + \frac{m_{2 \perp}}{\sqrt{s}} \exp( y_{4})\,, \quad
x_2 = \frac{m_{1 \perp}}{\sqrt{s}} \exp(-y_{3}) + \frac{m_{2 \perp}}{\sqrt{s}} \exp(-y_{4})\,,\nonumber\\
&&m_{1\perp} = \sqrt{m_{\tilde{\pi}^{0}}^{2} + p_{t,\tilde{\pi}^{0}}^{2}}\,, \quad
m_{2\perp} = p_{t,\tilde{\pi}^{0}}\,.
\end{eqnarray}
In this approach, transverse momenta of $\tilde{\pi}^{0}$ and outgoing $q/\bar{q}$ are strictly balanced.
Here we have introduced effective parton distribution which we define as:
\begin{equation}
q_{eff}(x,\mu^2) = \sum_f e_f^2 
\left( q_f(x,\mu^2) + {\bar{q}_{f}}(x,\mu^2) \right) \,,
\label{effective_quark_distribution}
\end{equation}
where we take $\mu^2 = p_{t,\tilde{\pi}^{0}}^2$.

The matrix element for the $\gamma q \to \tilde{\pi}^{0} q$ process
including masses of quarks reads:
\begin{eqnarray}
{\cal M}_{\gamma q \to \tilde{\pi}^{0} q}
= F_{\gamma \gamma}\,
\varepsilon^{\mu \nu \kappa \alpha} p_{1 \mu} p_{3 \nu} \,
\varepsilon^{(\gamma)}_\kappa(p_{1},\lambda_{1})
\frac{-i g_{\alpha \beta}}{k^{2}}\,
\bar{u}(p_{4},\lambda_{4}) \gamma^{\beta} u(p_{2},\lambda_{2}) \,,
\label{amp_gamq_piq}
\end{eqnarray}
were $k^{2} = (p_{2}-p_{4})^{2} = (p_{1}-p_{3})^{2}$.
The matrix element squared can be written in terms of the Mandelstam variables as
\begin{eqnarray}
\overline{|{\cal M}_{\gamma q \to \tilde{\pi}^{0} q}|^2}
=&& \frac{1}{4}\,F_{\gamma \gamma}^{2}\,
\frac{2e^{2}}{\hat{t}^{2}}\,
\left.\Bigl[ 2 (\hat{s}-m_{q}^{2})(k \cdot p_{1})(k \cdot p_{4})
           + 2 (m_{q}^{2}-\hat{u})(k \cdot p_{1})(k \cdot p_{2})
\right.\nonumber \\
&&\left.-4 m_{q}^{2} (k \cdot p_{1})^{2} +
\hat{t}(m_{q}^{2}-\hat{s})(m_{q}^{2}-\hat{u})
\right.\Bigr] \,,\\
\overline{|{\cal M}_{q \gamma \to \tilde{\pi}^{0} q}|^2}
=&& \frac{1}{4}\,F_{\gamma \gamma}^{2}\,
\frac{2e^{2}}{\hat{t}^{2}}\,
\left.\Bigl[ 2 (\hat{s}-m_{q}^{2})(k \cdot p_{2})(k \cdot p_{4})
           + 2 (m_{q}^{2}-\hat{u})(k \cdot p_{2})(k \cdot p_{1})
\right.\nonumber \\
&&\left.-4 m_{q}^{2} (k \cdot p_{2})^{2} +
\hat{t}(m_{q}^{2}-\hat{s})(m_{q}^{2}-\hat{u})
\right.\Bigr] \,.
\label{2to2_mq}
\end{eqnarray}
%

\subsection{$2 \to 3$ subprocess}
\label{sec:2to3_mechanism}

As the last contribution we discuss the diagram in Fig.~\ref{fig:diagrams_order3}.
The cross section for the partonic $q q' \to q {\tilde \pi}^0 q'$ 
process can be written as:
\begin{equation}
\sigma_{q q' \to q {\tilde \pi}^0 q'} = \frac{1}{2 {\hat s}}
\overline{ | {\cal M}_{q q' \to q {\tilde \pi}^0 q'}|^2 } {\cal J}
d \xi_1 d \xi_2 d y_{\tilde{\pi}^{0}} d \phi_{12} \,,
\end{equation}
where $\phi_{12}$ is the relative azimuthal angle between $q$ and $q'$,
$\xi_1 = \log_{10}(p_{1t}/1{\rm GeV})$ and $\xi_2 = \log_{10}(p_{2t}/1{\rm GeV})$,
where $p_{1t}$ and $p_{2t}$ are transverse momenta of outgoing $q$ and $q'$, respectively.

The matrix element for the $2 \to 3$ subprocess was calculated as:
\begin{eqnarray}
&&{\cal M}_{q q' \to q {\tilde \pi}^0 q'}
(\lambda_1, \lambda_2, \lambda_3, \lambda_4) = e^{2}\,
\bar{u}(p_{3},\lambda_{3}) \gamma^{\mu} u(p_{1},\lambda_{1}) \,
\frac{-i g_{\mu \nu}}{\hat{t}_{1}}\nonumber\\
&&\qquad \qquad \qquad \times 
\varepsilon^{\nu \nu' \alpha \beta} q_{1\alpha} q_{2\beta}\,
F_{\gamma\gamma}
\frac{-i g_{\nu' \mu'}}{\hat{t}_{2}}\,
\bar{u}(p_{4},\lambda_{4}) \gamma^{\mu'} u(p_{2},\lambda_{2})\,.
\label{ME_2to3}
\end{eqnarray}
For comparison, we shall also calculate the matrix element in the high-energy approximation:
\begin{eqnarray}
\bar{u}(p', \lambda') \,
\gamma^{\mu} \,
u(p, \lambda) 
\rightarrow 
(p' + p)^{\mu}\, 
\delta_{\lambda' \lambda}\,,
\label{ME_2to3_he}
\end{eqnarray}
often used in the literature in different context, see e.g. \ref{Lebiedowicz:2015cea}.
We have also obtained a formula for matrix element squared and checked
that it gives the same result as the calculation with explicit
use of spinors.

The total (phase-space integrated) cross section for technipion
production could be alternatively calculated as:
\begin{eqnarray}
\sigma_{p p \to {\tilde \pi}^0 jj} =
\int dx_1 dx_2 \sum_{f_1,f_2} q_{f_1}(x_1,\mu_{F}^2) q_{f_2}(x_2,\mu_{F}^2)
{\hat \sigma}_{f_1 f_2 \to f_1 {\tilde \pi}^0 f_2}({\hat s}) \,. 
\label{convolution1}
\end{eqnarray}
Limiting to $\gamma \gamma$-fusion processes only one can write:
\begin{eqnarray}
\sigma_{p p \to {\tilde \pi}^0 jj}^{\gamma \gamma} =
\int dx_{1} dx_{2} q_{eff}(x_1,\mu_{F}^2) q_{eff}(x_2,\mu_{F}^2) 
{\hat \sigma}^{eff}_{q q \to q \tilde{\pi}^{0} q}({\hat s}) \,,
\end{eqnarray}
%
%
%
where ${\hat \sigma}^{eff}_{q q \to q \tilde{\pi}^{0} q}({\hat s})$ is then the integrated cross section for the 
$f_1 f_2 \to f_1 {\tilde \pi}^0 f_2$ subprocess with both fractional quark/antiquark charges set to unity.

Above formula (\ref{convolution1}) is not very efficient when calculating
subprocess energy distribution. A more useful formula is:
\begin{equation}
\sigma_{p p \to \tilde{\pi}^0 jj} = 
\int dW \left( \hat{\sigma}^{eff}_{q q \to q \tilde{\pi}^{0} q}(W) 
\int d x_d \left( {\cal J} q_{eff}(x_1,\mu_{F}^2) q_{eff}(x_2,\mu_{F}^2) \right)
\right)\,.
\label{convolution2}
\end{equation}
Above $W = \sqrt{\hat{s}}$, $x_{d} = x_1 - x_2$, and ${\cal J}$ is a Jacobian of the transformation
from ($x_1$, $x_2$) to ($W$, $x_d$). In practical calculation first partonic 
$\hat{\sigma}^{eff}_{q q' \to q {\tilde \pi}^0 q'}$ (assuming elementary charges 
of quarks/antiquarks) is calculated as a function of $W$ 
on a grid and then the convolution with parton 
distributions is done as shown in Eq.~(\ref{convolution2}).
In the $2 \to 3$ hadronic calculations we take $\mu_{F}^2 = m_{\tilde{\pi}^0}^{2}$.

\section{Leading order VTC technipion signal in the diphoton channel}

\begin{figure}[!ht]
\includegraphics[width=0.3\textwidth]{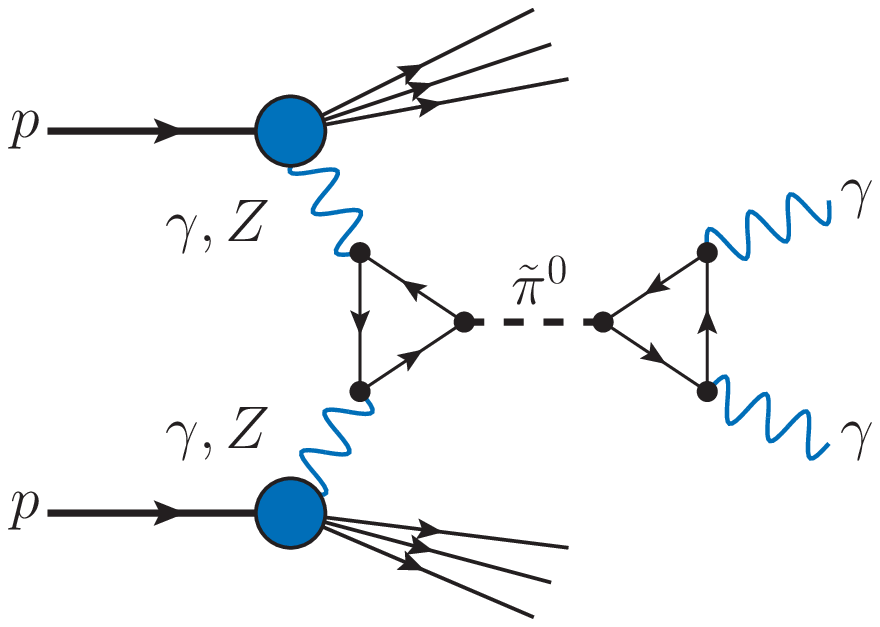}
\includegraphics[width=0.3\textwidth]{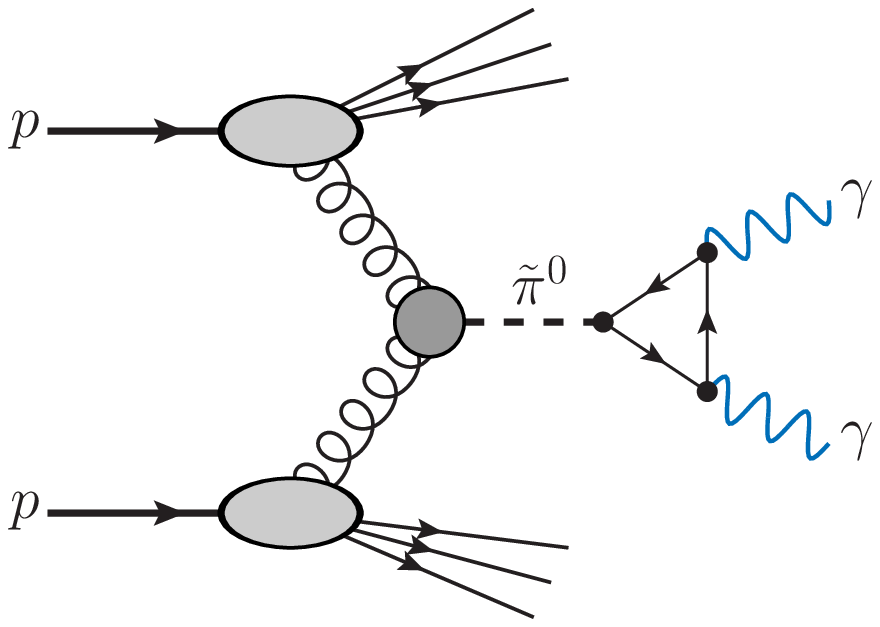}
  \caption{\label{fig:diagrams_gammagamma_fstate}
  \small
Leading order technipion signal in the diphoton channel in proton-proton collisions.
}
\end{figure}

In the case of VTC technipion model \cite{Pasechnik:2013bxa}, 
the amplitude for the $\gamma \gamma \to {\tilde \pi}^0 \to \gamma \gamma$ subprocess reads:
\begin{eqnarray}
&&{\cal M}_{\gamma\gamma \to {\tilde \pi}^0 \to \gamma\gamma}
(\lambda_1, \lambda_2, \lambda_3, \lambda_4) =
(\varepsilon^{(\gamma)\mu_{3}}(p_{3},\lambda_{3}))^{*}  
(\varepsilon^{(\gamma)\mu_{4}}(p_{4},\lambda_{4}))^{*}\,
\nonumber\\
&&\qquad \qquad \qquad 
\times 
\epsilon_{\mu_{3}\mu_{4} \nu_{3}\nu_{4}} p_{3}^{\nu_{3}} p_{4}^{\nu_{4}}\,
F_{\gamma\gamma}\, 
\frac{i}{\hat{s}-m_{\tilde{\pi}^{0}}^{2}+i m_{\tilde{\pi}^{0}}\Gamma_{tot}}
\nonumber\\
&&\qquad \qquad \qquad 
\times 
\epsilon_{\mu_{1}\mu_{2} \nu_{1}\nu_{2}} p_{1}^{\nu_{1}} p_{2}^{\nu_{2}}\,
F_{\gamma\gamma}\,
\varepsilon^{(\gamma)\mu_{1}}(p_{1},\lambda_{1})
\varepsilon^{(\gamma)\mu_{2}}(p_{2},\lambda_{2})\,.
\label{amp_signal}
\end{eqnarray}
The $\Gamma_{tot}$ can be calculated from a model or taken from recent experimental data. In the following we take
the calculated value of $\Gamma_{tot}$ and $m_{\tilde{\pi}^{0}}$ = 750 GeV.
The mass scale of the degenerate techniquarks $m_{\tilde Q}$ is in principle another free parameter (see e.g. Ref.~\cite{Lebiedowicz:2013fta}).

The cross section for the signal 
(see the left panel of Fig.~\ref{fig:diagrams_gammagamma_fstate})
is calculated as ($\mu_F^2 = p_{t,\gamma}^2$):
\begin{equation}
\frac{d \sigma}{d y_3 d y_4 d^2 p_{t,\gamma}}
= \frac{1}{16 \pi^2 {\hat s}^2} \sum_{ij}
x_1 \gamma^{(i)}(x_1,\mu_F^2) x_2 \gamma^{(j)}(x_2,\mu_F^2)
\overline{|
{\cal M}_{\gamma\gamma \to \tilde{\pi}^{0} \to \gamma\gamma} |^2} \,,
\label{inclusive_gamgam_signal}
\end{equation}
where 
\begin{eqnarray}
x_1 = \frac{p_{t,\gamma}}{\sqrt{s}} \left[ \exp({y_{3}}) + \exp(y_{4}) \right]\,, \quad
x_2 = \frac{p_{t,\gamma}}{\sqrt{s}} \left[ \exp({-y_{3}}) + \exp(-y_{4}) \right]\,.
\end{eqnarray}
%
\section{One-family walking technicolor model}

In the one-family walking technipion model discussed recently in
Ref.~\cite{Matsuzaki:2015che} (see also references therein) 
the partial $gg$ and $\gamma \gamma$ decay widths 
are given as:
\begin{eqnarray}
\Gamma(P^0 \to gg) &=& 
\frac{N_{TC}^2 \,\alpha_{s}^{2} \,G_F \,m_{P^{0}}^3}{12 \sqrt{2} \pi^3}\,,
\label{partial_decay_widths_gg}\\
\Gamma(P^0 \to \gamma \gamma) &=& 
\frac{N_{TC}^2 \,\alpha_{em}^{2}\,G_F \,m_{P^{0}}^3}{54 \sqrt{2} \pi^3}\,,
\label{partial_decay_widths_gamgam}
\end{eqnarray}
where $\alpha_{s} \equiv g_{s}^{2}/(4\pi)$ is the strong coupling constant,
$N_{TC}$ is the number of technicolors in the walking technicolor model.
For $N_{TC}$ = 3 we get:
\begin{center}
$\Gamma(P^0 \to gg) =$ 1.2 GeV, \qquad 
$\Gamma(P^0 \to \gamma \gamma) =$ 1.2 MeV.
\end{center}
The decay into two gluons is in this model 
the dominant decay channel \cite{Matsuzaki:2015che}.
The total decay width in the model is therefore also much smaller
than the 45 GeV reported in Refs~\cite{ATLAS:2015,CMS:2015dxe} and used in many
very recent analyses. We will return to this point in the result section.
It is interesting that the model gives roughly correct size of the
signal for $N_{TC}$ = 3, 4, without any additional tuning.

The cross section for the signal in this scenario is calculated then as ($\mu_F^2 = p_{t,\gamma}^2$):
\begin{equation}
\frac{d \sigma}{d y_3 d y_4 d^2 p_{t,\gamma}}
= \frac{1}{16 \pi^2 {\hat s}^2}
\frac{1}{N_c^2-1} x_1 g(x_1,\mu_F^2) x_2 g(x_2,\mu_F^2)
\overline{|
{\cal M}_{g g \to {\tilde \pi}^0 \to \gamma\gamma} |^2} \,.
\label{exclusive_gg_signal}
\end{equation}

The color factor $N_c$ guarantees that the technipion resonance is a QCD-white object. 
Clearly, in this model the gluon-gluon fusion is the dominant reaction
mechanisms. This would also mean a rather large cross section for 
the dijet production of the order of a few pb. We shall show below
whether this is compatible with the existing data for dijets production.

\section{Production mechanisms of background in $\gamma\gamma$ channel}

\begin{figure}[!ht]
(a)\includegraphics[width=0.22\textwidth]{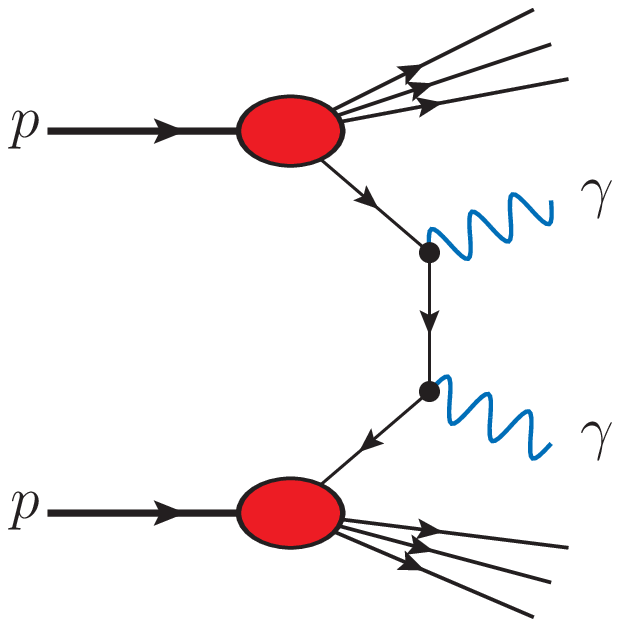}
(b)\includegraphics[width=0.26\textwidth]{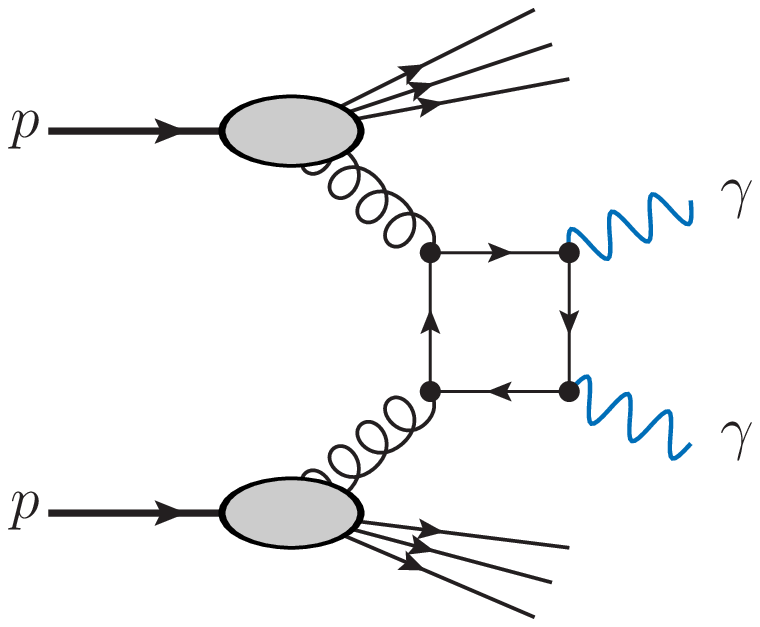}
(c)\includegraphics[width=0.225\textwidth]{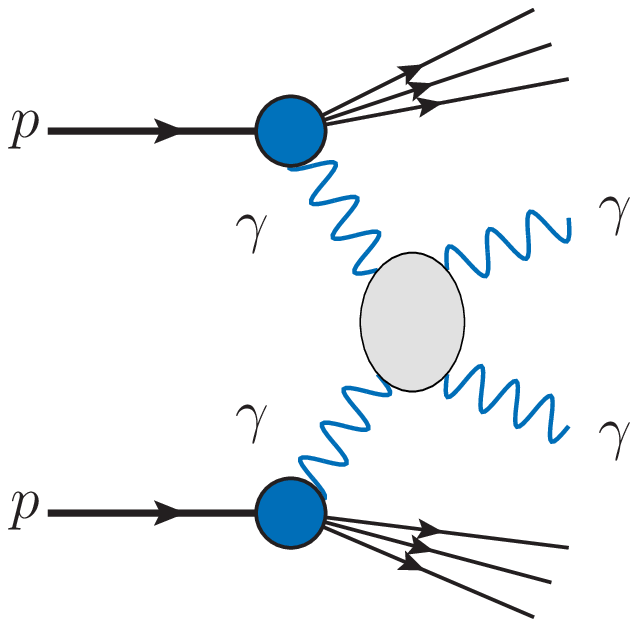}
  \caption{\label{fig:diagrams_background}
  \small
Background mechanisms of $\gamma\gamma$ pairs production in proton-proton collisions.
}
\end{figure}
In the present exploratory analysis we consider the background contributions
shown in Fig.~\ref{fig:diagrams_background}. These include
the $q \bar q$ annihilation (diagram (a)), 
the gluon-gluon fusion via quark boxes (diagram (b)), 
and the photon-photon fusion via lepton, quark and $W$-boson loops (diagram (c)).

They were found recently to be very important for $W^+ W^-$ production
with large  $M_{W^{+}W^{-}}$ \cite{Luszczak:2014mta}. 
In addition, this type of background could interfere with the signal. 
For simplicity we shall neglect these interference effects in the present paper.

\subsection{Background $q \bar q$ annihilation contribution}

The lowest order process for diphoton production is quark-antiquark annihilation. 
The cross section for $q \bar{q}$ annihilation can be written as: 
\begin{equation}
\frac{d \sigma}{d y_3 d y_4 d^2 p_{t,\gamma}} = 
\frac{1}{16 \pi^2 {\hat s}^2} 
\sum_f x_1 q_f(x_1, \mu_F^2) x_2 \bar q_f(x_2,\mu_F^2) 
\overline{| {\cal M}_{q \bar q \to \gamma\gamma} |^2} \,.
\label{inclusive_qqbar}
\end{equation}
The formula for the matrix element squared for the 
$q \bar q \to \gamma \gamma$ subprocess 
can be found e.g. in Ref.~\cite{Berger:1983yi}.
In our calculation we include only three quark flavours ($u$, $d$, $s$).

\subsection{Background $gg$ fusion contribution}

For a test and for a comparison we also consider the gluon-gluon contribution to the inclusive cross section.
The photons produced in $pp(\bar{p}) \to \gamma\gamma + X$ are expected to be dominantly produced 
by the quark-antiquark annihilation ($q\bar{q} \to \gamma\gamma$) and by the gluon-gluon fusion 
($gg \to \gamma\gamma$) through a quark-box diagram. The latter process is important especially 
at low diphoton invariant masses in kinematic region with high gluon luminosity.

In the lowest order of pQCD the formula for inclusive cross section can be written as
\begin{equation}
\frac{d \sigma}{d y_3 d y_4 d^2 p_{t,\gamma}} = \frac{1}{16 \pi^2
{\hat s}^2} x_1 g(x_1, \mu_F^2) x_2 g(x_2,\mu_F^2) \overline{| {\cal
M}_{gg \to \gamma\gamma} |^2} \,.
\label{inclusive_gg}
\end{equation}
The corresponding matrix elements have been discussed in detail e.g. in Ref.~\cite{Glover:1988fe}.

\subsection{Background $\gamma\gamma$ fusion contribution}

The cross section of $\gamma\gamma$ production via $\gamma \gamma$ fusion in $pp$
collisions can be calculated in the same way as in the parton model
in the so-called equivalent photon approximation as
\begin{equation}
\frac{d \sigma}{d y_3 d y_4 d^2 p_{t,\gamma}}
= \frac{1}{16 \pi^2 {\hat s}^2} 
\sum_{ij} x_1 \gamma^{(i)}(x_1, \mu_F^2) x_2 \gamma^{(j)}(x_2, \mu_F^2)
\overline{|{\cal M}_{\gamma\gamma \to \gamma\gamma} |^2} \,.
\label{inclusive_gamgam}
\end{equation}

In practical calculations for elastic fluxes we shall use parametrization proposed in Ref.~\cite{Drees:1994zx}. 
The loop-induced helicity matrix element for the $\gamma \gamma \to \gamma \gamma$ subprocess 
was calculated by using the {\tt Mathematica} package {\tt FormCalc} \cite{Hahn:1998yk} 
and the {\tt LoopTools} library based on \cite{vanOldenborgh:1989wn} 
to evaluate one-loop integrals. 
In numerical calculations we include box diagrams with
leptons, quarks as well as with $W$ bosons. 
At high diphoton invariant masses the inclusion of diagrams with 
$W$ bosons in loops is crucial, see e.g. \cite{Lebiedowicz:2013fta}. 

\section{Discussion of results and consistency checks with existing experiments}
\label{sec:Results}

\subsection{Signal of technipion}

Let us first summarize integrated cross sections for the VTC scenario.
In Table~I we have collected cross sections for different QED orders.
For consistency all cross sections were calculated with the MRST(QED) parton distributions \cite{Martin:2004dh}.
In this calculation we have used $g_{TC} = 10$ for example and our benchmark parameters.
Surprisingly, different contributions are of the same order of magnitude. 
Note that with $g_{TC} = 10$ we get the cross section of correct order of magnitude.
To describe the experimental signal more precisely $g_{TC}$ can be
rescalled. A result consistent with the cross section extracted from
experimental ATLAS and/or CMS data is obtained with $g_{TC}$ = 20.
\begin{table}
{\caption
{\small Hadronic cross section in fb for neutral
technipion production at $\sqrt{s}$ = 1.96, 7, 13, 100 TeV
for different contributions 
shown in Figs.~\ref{fig:diagrams_order1} - \ref{fig:diagrams_order3}.
Here we assume $g_{TC}$ = 10 and $m_{{\tilde Q}} = 0.75 \,m_{{\tilde \pi}^{0}}$.}}
\label{tab:table1}
\begin{tabular}{|l|c|c|c|c|}
\hline
Component                           &1.96 TeV &7 TeV  &13 TeV &100 TeV   \\
\hline
2 $\to$ 1 (in, in)                & 1.37E-3        & 0.16      & 0.55 &  \;\;8.08       \\
2 $\to$ 1 (in, el)                & 0.22E-3        & 0.05      & 0.15 &  \;\;1.88       \\
2 $\to$ 1 (el, in)                & 0.22E-3        & 0.05      & 0.15 &  \;\;1.88       \\
2 $\to$ 1 (el, el)                & 0.03E-3        & 0.01      & 0.04 &  \;\;0.42       \\
\textbf{2 $\to$ 1,  sum of all}  & \textbf{1.84E-3}        & \textbf{0.27}      & \textbf{0.89} &     \textbf{12.26}       \\
\hline
2 $\to$ 2 (in, in), two diagrams          & 0.74E-3 & 0.14 & 0.49 & \;\;7.69  \\
2 $\to$ 2 (in, el) and (el, in)            & 0.13E-3 & 0.05 & 0.19 & \;\;2.93  \\
\textbf{2 $\to$ 2,  sum of all}          & \textbf{0.87E-3} & \textbf{0.19} & \textbf{0.68} &    \textbf{10.62}  \\
2 $\to$ 2,  sum of all, $p_{t,jet} >$ 10 GeV&         &       & 0.43 & \;\;8.03       \\
2 $\to$ 2,  sum of all, $p_{t,jet} >$ 20 GeV&         &       & 0.35 & \;\;6.99       \\
2 $\to$ 2,  sum of all, $p_{t,jet} >$ 50 GeV&         &       & 0.25 & \;\;5.42       \\
\hline
\textbf{2 $\to$ 3}                           &   \textbf{0.14E-3}      & \textbf{0.09}      & \textbf{0.46}  &  \textbf{16.71}   \\
2 $\to$ 3,  $p_{t,jet} >$ 10 GeV             &         &       & 0.04  &  \;\;1.41     \\
\hline
\end{tabular}
\end{table}

We shall discuss now some specific results for different processes.

Let us start from the signal in the VTC model \cite{Pasechnik:2013bxa}. 
In our calculations here we use MRST04(QED) parton distributions
\cite{Martin:2004dh}. In this ($2 \to 1$) calculation the $\sigma^{({\rm in,\, in})}/\sigma_{tot} \approx 0.6$.
For comparison  $\sigma^{({\rm el, \,el})}/\sigma_{tot} \approx 0.04$. 
This means that for the purely exclusive reactions we get 
$\sigma_{p p \to p p \gamma \gamma}^{{\rm signal}} \lesssim 0.2$~fb. 
In order to get experimental value in the fiducial volume 
$\sigma_{pp \to \gamma \gamma} \approx 7$~fb we have to assume a rather large value of $g_{TC} \approx 40$. 
This is a huge value and puts into doubts perturbative approach. 
We will not worry here about the conceptual problem
and test further consequences. The corresponding $\Gamma_{tot} =
0.16$~GeV (very narrow width scenario).
The narrow width approximation was preferred by the CMS analysis \cite{CMS:2015dxe}.

As discussed in section \ref{sec:2to3_mechanism} the production of technipion can be calculated also
as $2 \to 3$ subprocess with intermediate (off-shell) photons.
We neglect here intermediate $Z$ exchanges for simplicity.
We shall discuss now how such results compare to the previous results 
obtained in the $2 \to 1$ ($\gamma \gamma \to \tilde{\pi}^0$) 
or $2 \to 2$ ($\gamma \gamma \to \tilde{\pi}^0 \gamma \gamma$) 
when photons from the decay of neutral technipion 
are considered, and ``initial'' photons are assumed to be on-shell.

\begin{figure}[!ht]
\includegraphics[width=0.4\textwidth]{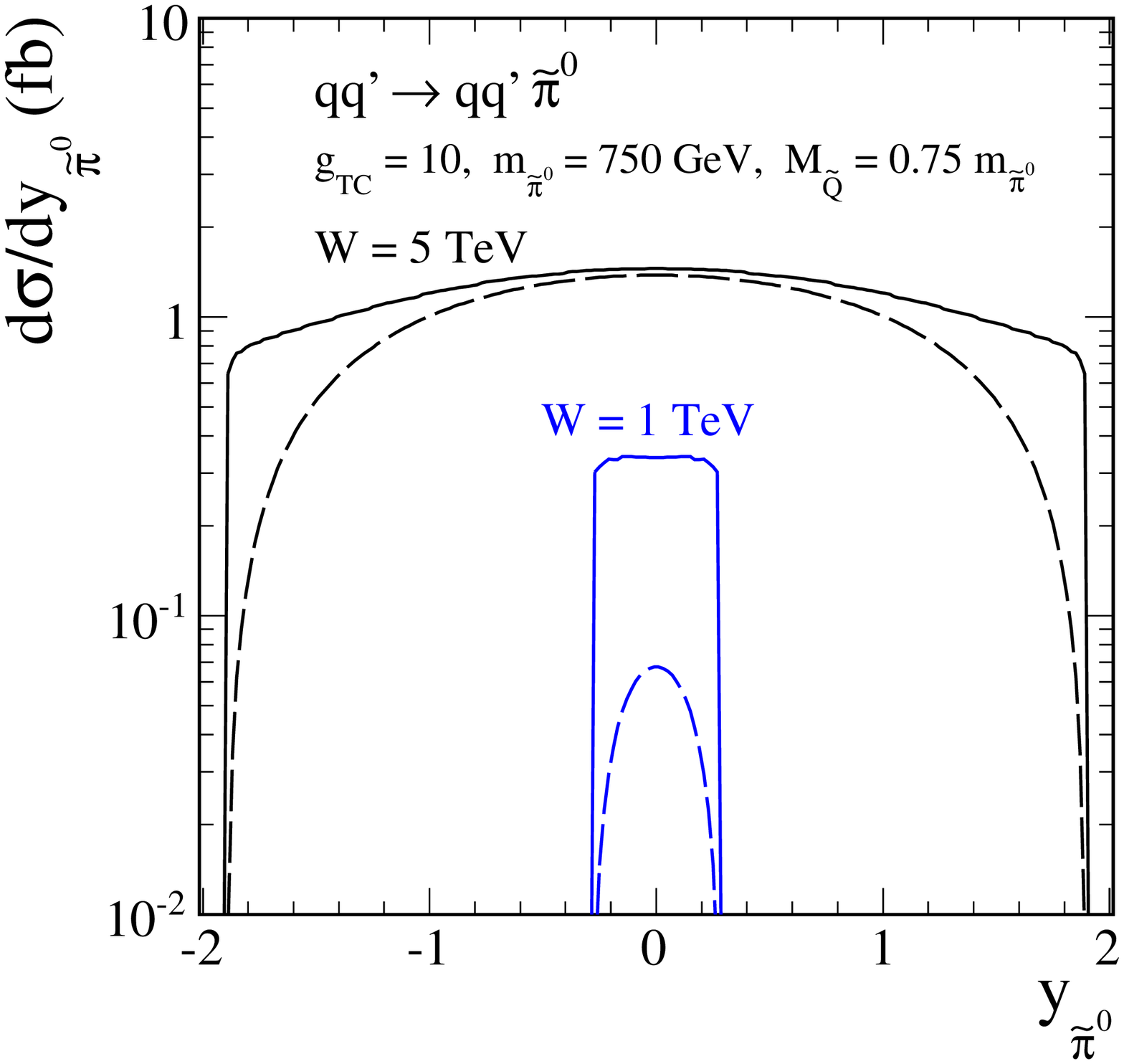}
\includegraphics[width=0.4\textwidth]{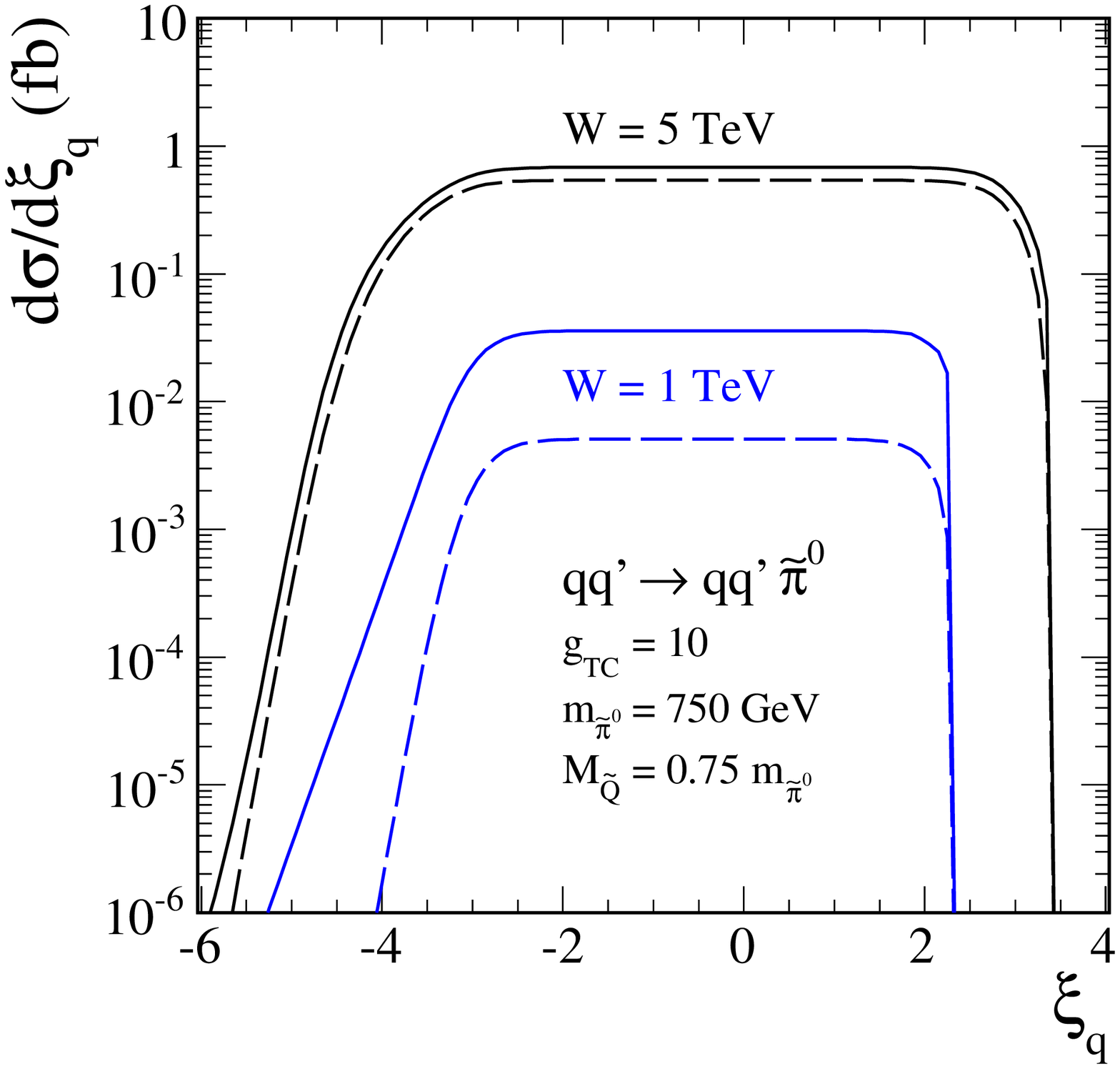}
\includegraphics[width=0.4\textwidth]{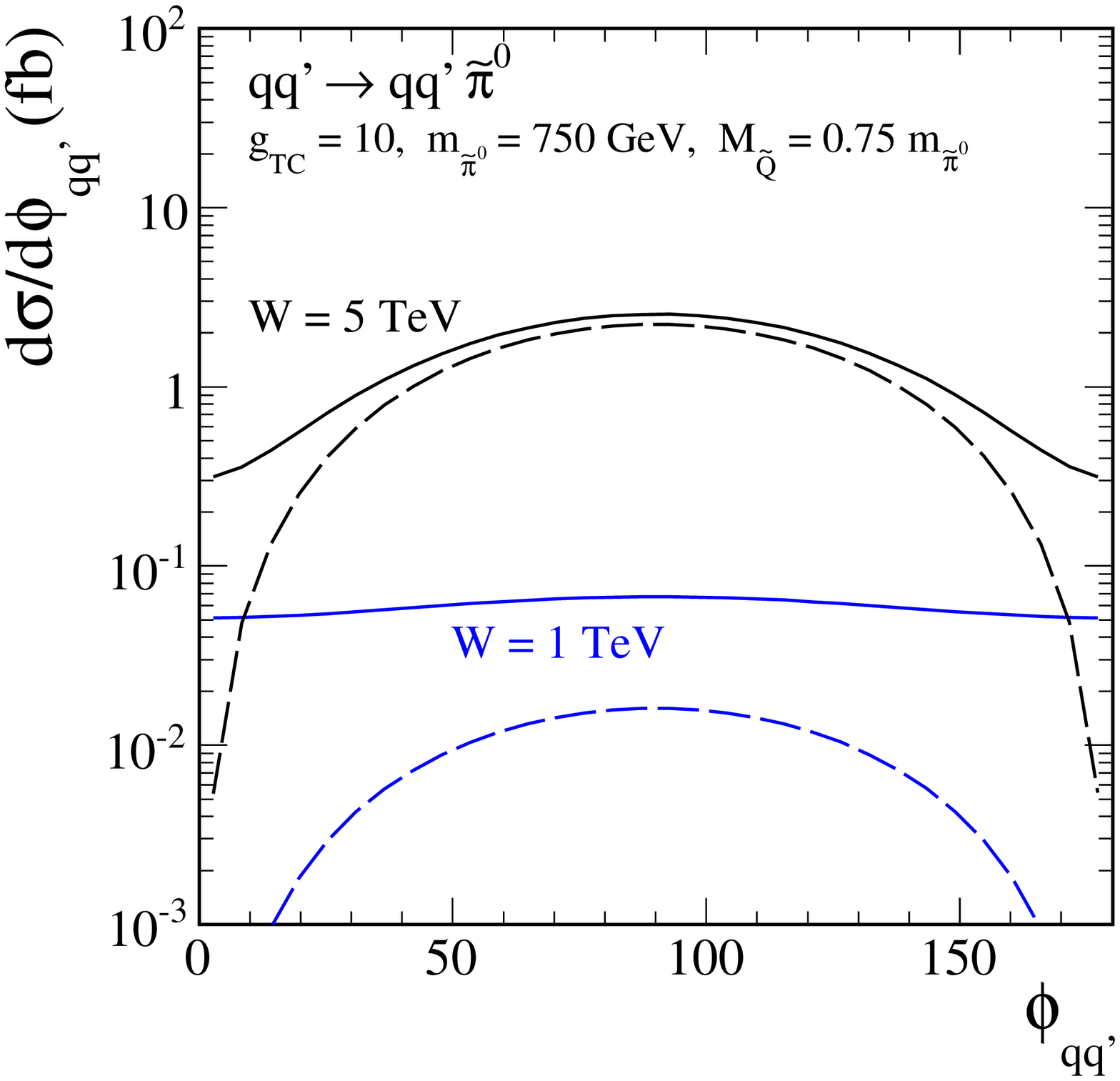}
  \caption{\label{fig:partonic_2to3}
  \small
Differential $q q' \to q \tilde{\pi}^0 q'$ cross sections
(assuming unit charges)
for two subprocess energies $W$ = 1 TeV (blue lower lines) 
and $W$ = 5 TeV (black upper lines).
In this calculation we use the set of parameters as shown in the legend.
The dashed lines were obtained in the high-energy approximation
while the solid lines represent the exact (with spinors) calculations.
}
\end{figure}

In Fig.~\ref{fig:partonic_2to3} we show the $\tilde{\pi}^0$
rapidity distribution.
We show here both the contribution of $2 \to 1$ (see Fig.~\ref{fig:diagrams_order1}) 
and two contributions of $2 \to 2$ (see Fig.~\ref{fig:diagrams_order2}) 
processes. Each of the $2 \to 2$ ($\gamma q \to {\tilde \pi}^0 q$ and $q \gamma \to {\tilde \pi}^0 q$) contributions separately is asymmetric
with respect to $y_{{\tilde \pi}^0}$ = 0. The sum is then similar as for the $2 \to 1$
contribution. The calculation of rapidity distribution of ${\tilde \pi}^0$ 
for the mechanism with $2 \to 3$ subprocess (see Fig.~\ref{fig:diagrams_order3})
is more complicated and will be omitted here.

\begin{figure}[!ht]
\includegraphics[width=0.4\textwidth]{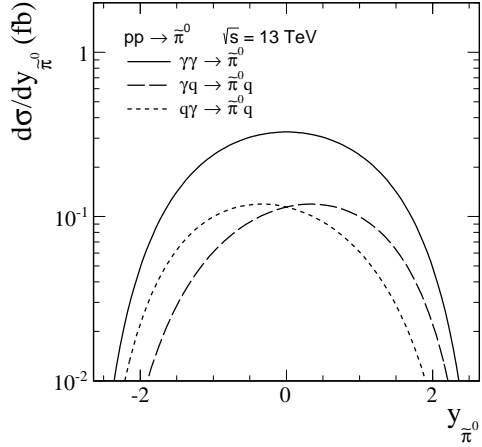}
  \caption{\label{fig:partonic_all}
  \small
Distribution in rapidity of neutral technipion
for all partonic subprocesses for the $p p \to \tilde{\pi}^0$ at $\sqrt{s}$ = 13 TeV.
In the calculation we take 
$g_{TC} = 10$, $m_{\tilde{\pi}^0} = 750$~GeV, $m_q$ = 1 MeV (for all flavours),
$m_{\tilde Q} = 0.75 \,m_{\tilde{\pi}^0}$ (for both techni-flavours).
}
\end{figure}

\begin{figure}[!ht]
\includegraphics[width=0.35\textwidth]{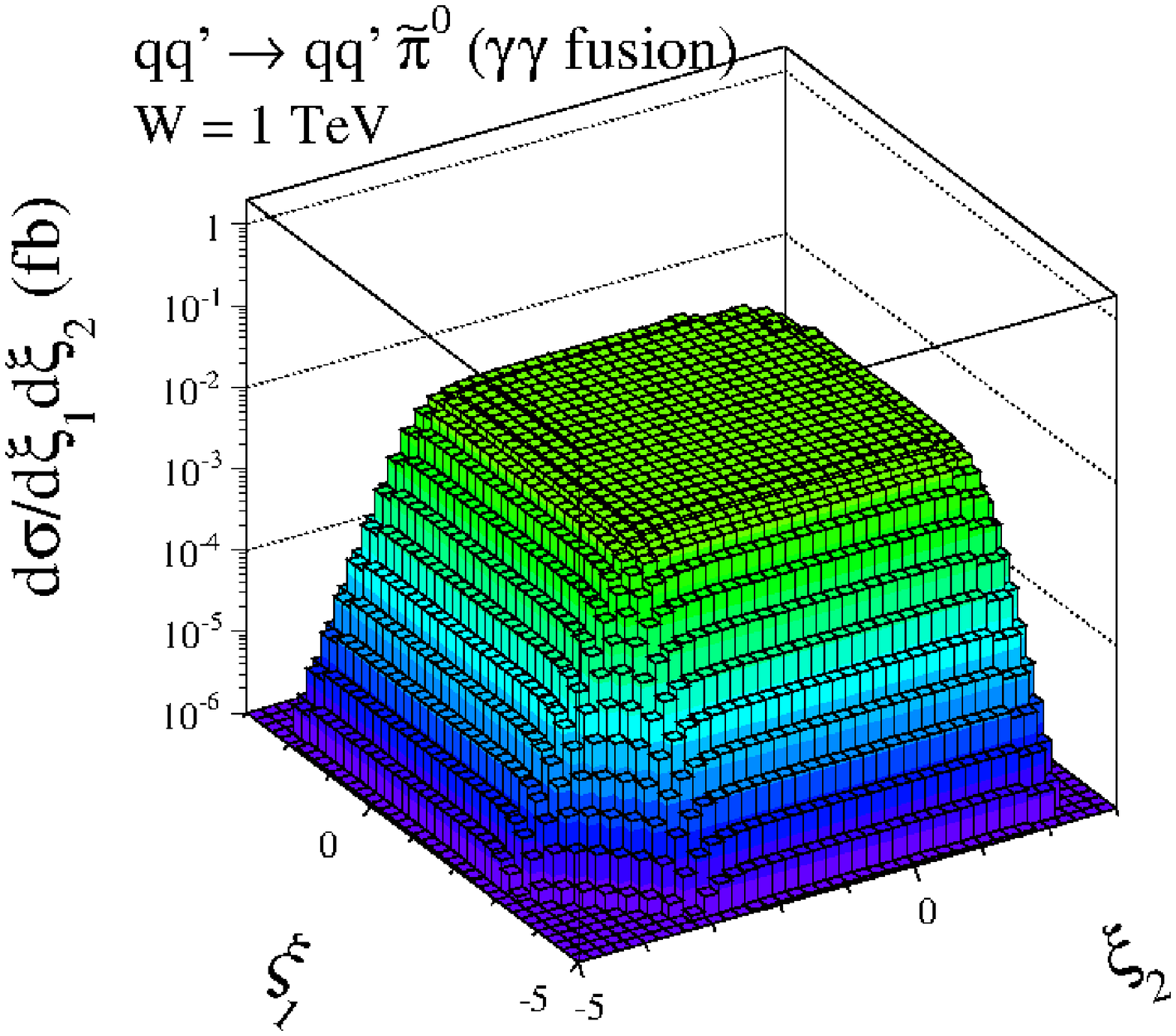}
\includegraphics[width=0.35\textwidth]{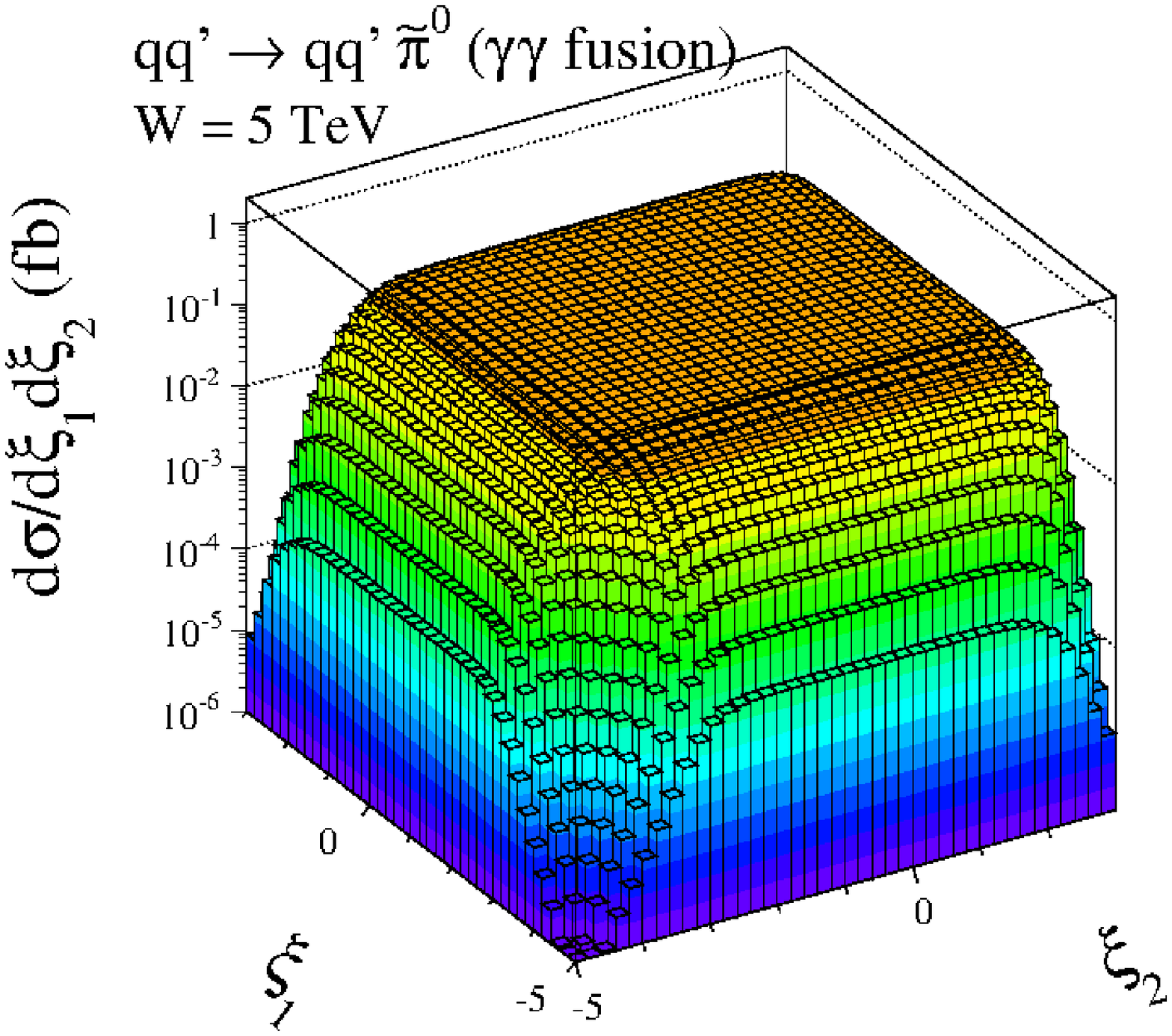}
  \caption{\label{fig:partonic_2to3_xi1xi2}
  \small
Differential $q q' \to q {\tilde \pi}^0 q'$ cross section (unit charges) 
as a function of $\xi_1 = log_{10}(p_{1t}/1{\rm GeV})$ 
and $\xi_2 = log_{10}(p_{2t}/1{\rm GeV})$
for two subprocess energies: (a) W = 1 TeV (left panel) and (b) W = 5 TeV (right panel). 
In this calculation we use $g_{TC}$ = 10 for example.
}
\end{figure}

Before we shall present corresponding cross section(s) for 
the $p p \to \tilde{\pi}^0 j j$ reaction we wish to show some interesting results
for the partonic $q q' \to q \tilde{\pi}^0 q'$ cross section.
When discussing partonic cross section we shall show results for
unit electric charges of quarks/antiquarks. The real quark charges are then included in the
hadronic cross section. The integrated partonic cross section is an integral over four 
properly chosen kinematic variables. 
Different choices are possible a priori.
In the present calculations we used: $\xi_1 = log_{10}(p_{1t}/1{\rm GeV})$, $\xi_2 = log_{10}(p_{2t}/1{\rm GeV})$, 
rapidity of technipion and relative azimuthal angle between outgoing
quarks/antiquarks. Especially the distribution in $\xi_1$ and $\xi_2$
is very interesting and could be even surprising.
In Fig.~\ref{fig:partonic_2to3_xi1xi2} we show the two-dimensional
distribution in $(\xi_1,\xi_2)$. The distribution is surprisingly flat
over a broad range of $\xi_1$ and $\xi_2$ which justifies use of the
variables. Here, both very small $p_{1t},p_{2t} \ll$ 1 GeV and very large 
$p_{1t},p_{2t} \gg$ 10 GeV transverse momenta of quarks/antiquarks 
contribute. This would mean that a large fraction of the cross section is associated
with one or two jets. 
We shall return to this point in the conclusion section.

In Fig.~\ref{fig:partonic_2to3} we show 
three different distributions:
in rapidity of the technipion,
in the integration variable $\xi_1$ (or $\xi_2$), and in relative
azimuthal angle between the associated ``jets''.
Only some relative azimuthal directions between jets 
are preferred by the central $\gamma \gamma \to {\tilde \pi}^0$ vertex. 
This predictions could be also used in searches for technipion
associated with one or two jets, which we strongly advocate.

\begin{figure}[!ht]
\includegraphics[width=0.4\textwidth]{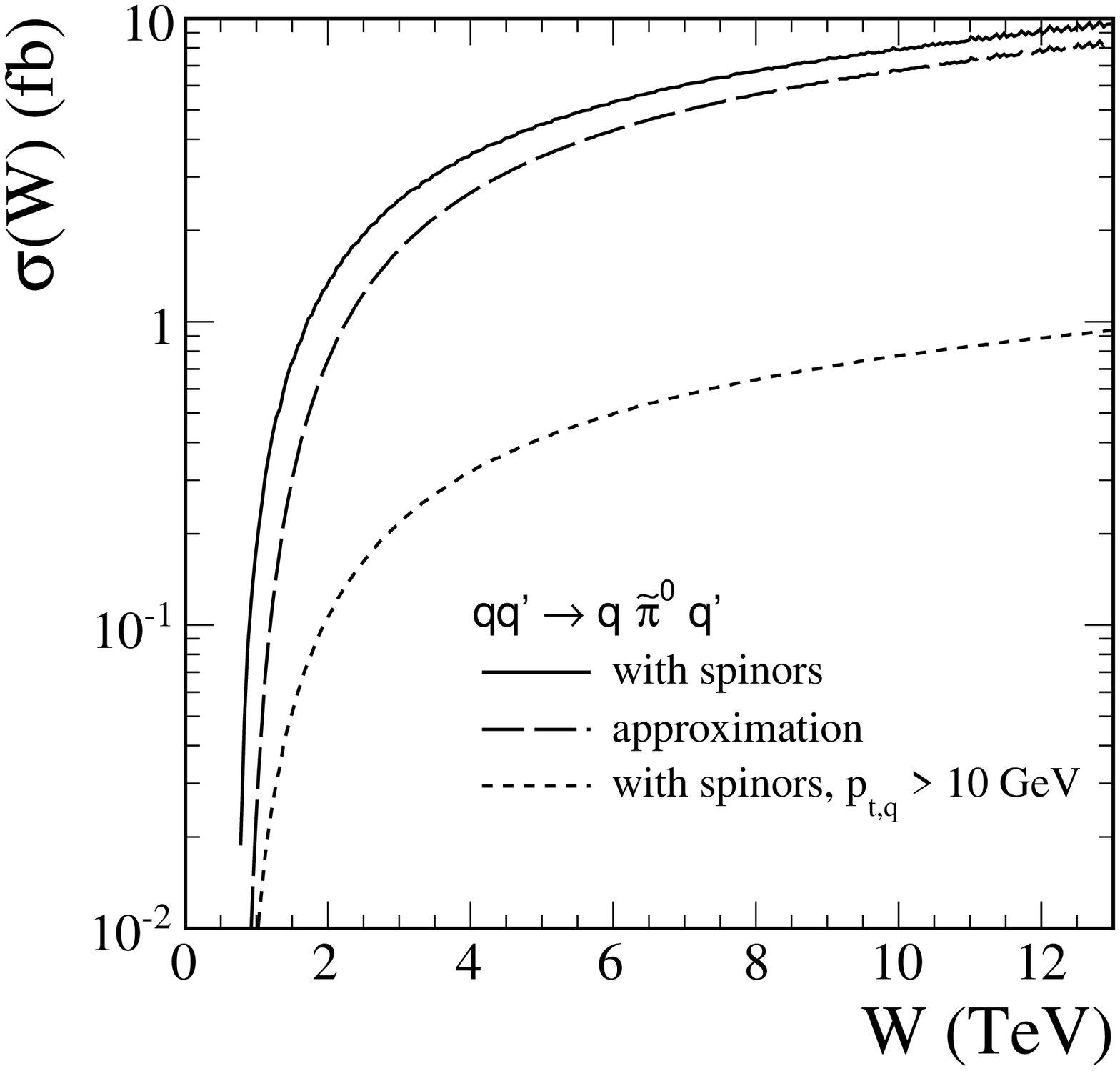}
\includegraphics[width=0.4\textwidth]{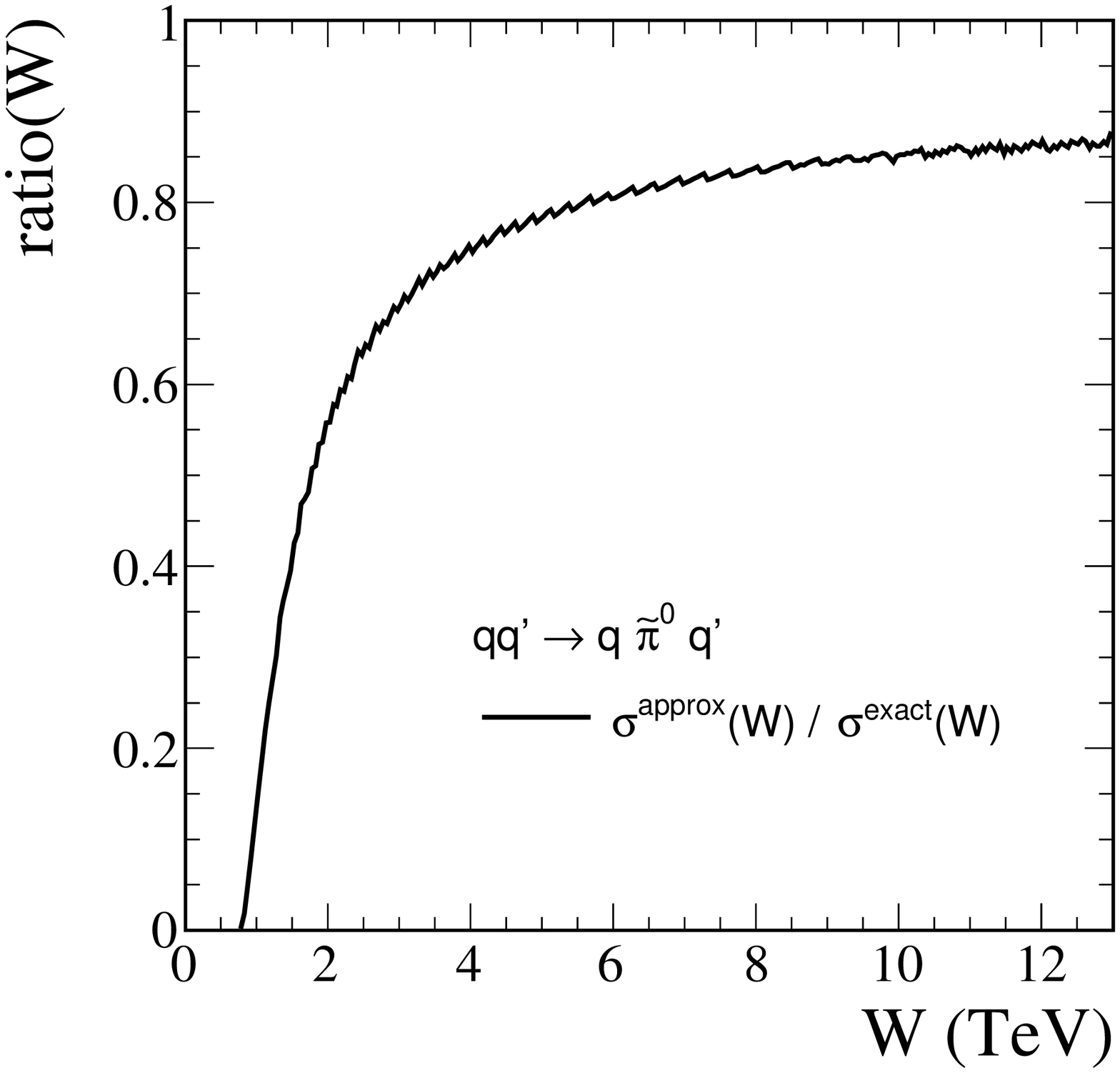}
  \caption{\label{fig:partonic_2to3_W}
  \small
Left panel: Total partonic $q q' \to q {\tilde \pi}^0 q'$
cross section (unit charges) as a function of subprocess energy.
In this calculation $g_{TC}$ = 10 was used for example. 
The solid line is for the calculation with
spinors while the dashed line was obtained in the high-energy approximation.
The dotted line corresponds to calculation for the exact case
with the cut on both transverse momenta of (anti)quarks
$p_{1t}, p_{2t} >$ 10 GeV.
Right panel: Ratio of the total partonic  $q q' \to q {\tilde \pi}^0 q'$ cross section
(unit charges) in high-energy approximation to the one with spinors
as a function of subprocess energy.
}
\end{figure}

The partonic cross section as a function of subprocess energy $W$
is shown in Fig.~\ref{fig:partonic_2to3_W} (left panel). We observe a quick rise
of the cross section from the threshold $W = \sqrt{{\hat s}} =  m_{\tilde \pi}$.
In addition to the result with fermions (i.e. including spinors)
we show result for high-energy approximation often made, e.g. in
diffractive processes (see e.g. Ref.~\cite{Lebiedowicz:2016ioh}). 
A huge difference between the two results can be observed especially close to the threshold.
We show also (dash-dotted line) the cross section when the cut 
on transverse momenta of (anti)quarks $p_{1t}, p_{2t} >$ 10 GeV
is imposed in addition. These cross sections are smaller by order of
magnitude than the total (without cuts) cross sections but relative
background contributions are expected to be smaller.
In the right panel of Fig.~\ref{fig:partonic_2to3_W}
we show the dependence on subprocess energy of the ratio of 
the partonic cross section obtained in the high-energy approximation
(which coincides with the result for spinless objects) 
to the one obtained with spinors. At high subprocess energy the two results start
to converge but the energy must be really large in order that the
approximation is really good.

\begin{figure}[!ht]
\includegraphics[width=0.4\textwidth]{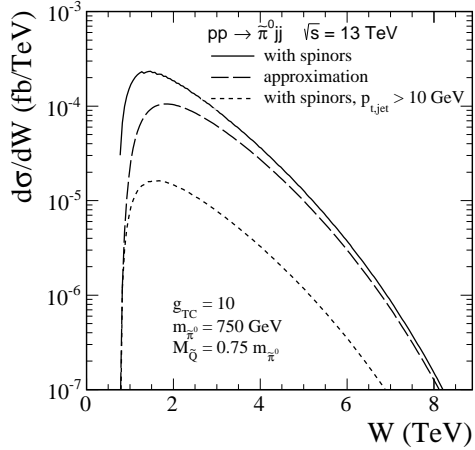}
  \caption{\label{fig:hadronic_2to3_W}
  \small
The distribution in the energy in the partonic subprocess
for the $p p \to {\tilde \pi}^0 jj$ at $\sqrt{s}$ = 13 TeV.
In this calculation we use $g_{TC}$ = 10 for example. 
The solid and dotted lines are for the calculation with
spinors while the dashed line was obtained in the high-energy approximation.
}
\end{figure}

Now we proceed to calculations of hadronic cross sections for
technipion production. In calculating hadronic cross section we use leading-order 
MSTW08 parton distributions \cite{Martin:2009iq}.
In Fig.~\ref{fig:hadronic_2to3_W} we show distribution in subsystem energy
in proton-proton collision for $\sqrt{s}$ = 13 TeV, corresponding to
the actual experiments performed by the ATLAS and CMS Collaborations.
Here we observe that close-to-threshold subenergies are crucial 
in the calculation. This is the region where the exact (with spinors) and approximate
(``spinless quarks'') results differ the most. 

Performing convolution of the partonic cross section with 
the quark/antiquark distributions we get for $g_{TC}$ = 10:
$\sigma$ = 0.46 fb for the exact case 
and $\sigma$ = 0.24 fb in the high-energy approximation.
Corresponding hadronic cross section with extra cuts on quark transverse
momenta $p_{1t}, p_{2t} >$ 10 GeV (for the exact case) 
is 0.04 fb, order of magnitude less than the full phase space one.
The cross section for calculation with spinors is only factor of
two larger than the one for fictitious spinless quarks,
which is often called in the literature high-energy approximation.
This is because both small and large $\sqrt{{\hat s}}$ regions enter into 
the calculation of the hadronic cross section.

\begin{figure}[!ht]
\includegraphics[width=0.6\textwidth]{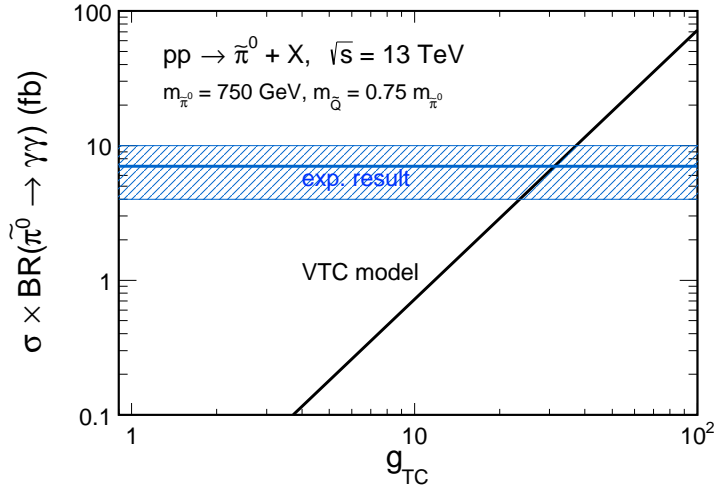}
  \caption{\label{fig:hadronic_g_TC}
  \small
The dependence of the hadronic $p p \to {\tilde \pi}^0 + X$ cross section 
on $g_{TC}$ together with crudely estimated by us experimental result 
at the LHC \cite{ATLAS:2015,CMS:2015dxe}. 
The solid black line represents our result for the technipion production in the VTC model.
}
\end{figure}

The dependence of the cross section on $g_{TC}$ is shown in
Fig~\ref{fig:hadronic_g_TC}. With $g_{TC}$ = 20 
we are at the ballpark with the ``measured'' value at $\sqrt{s} = 13$~TeV.
The value of $g_{TC}$ could be even smaller when exchange of $Z$ bosons 
is included. This will be done elsewhere.

\subsection{Comparison with background contributions}

Now we can look at differential distributions and compare the technipion signal 
to irreducible SM background contributions in the first (photon PDFs) approach.
The distributions in rapidity of photons and transverse momentum of 
one of them $p_{t,\gamma}$ can be calculated in a straightforward 
way from Eqs.~(\ref{inclusive_gamgam_signal}), (\ref{inclusive_qqbar}), (\ref{inclusive_gg}), 
(\ref{inclusive_gamgam}). 
In turn the distribution in diphoton invariant mass can be obtained by an appropriate binning.

\begin{figure}[!ht]
\includegraphics[width=0.3\textwidth]{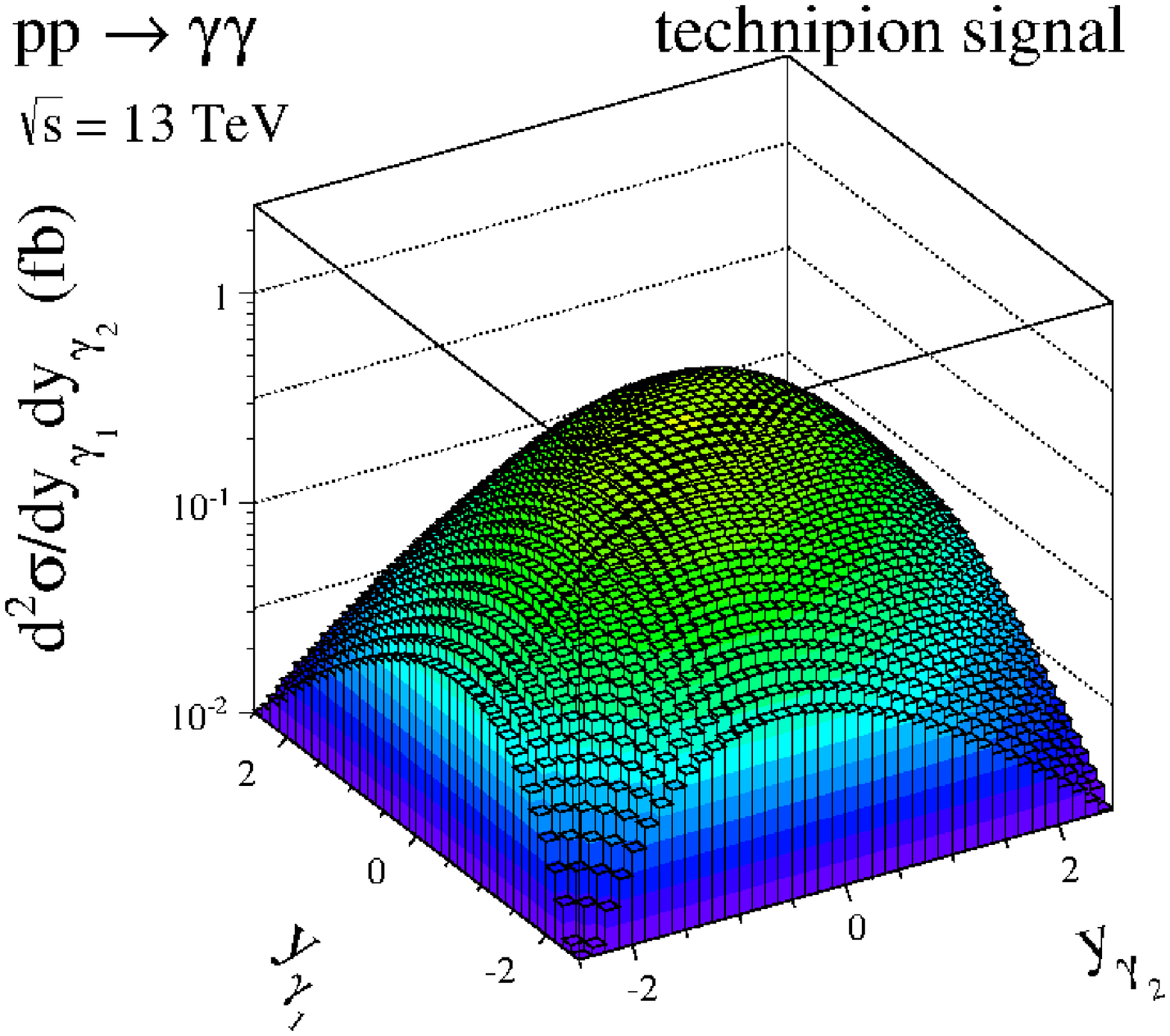}
\includegraphics[width=0.3\textwidth]{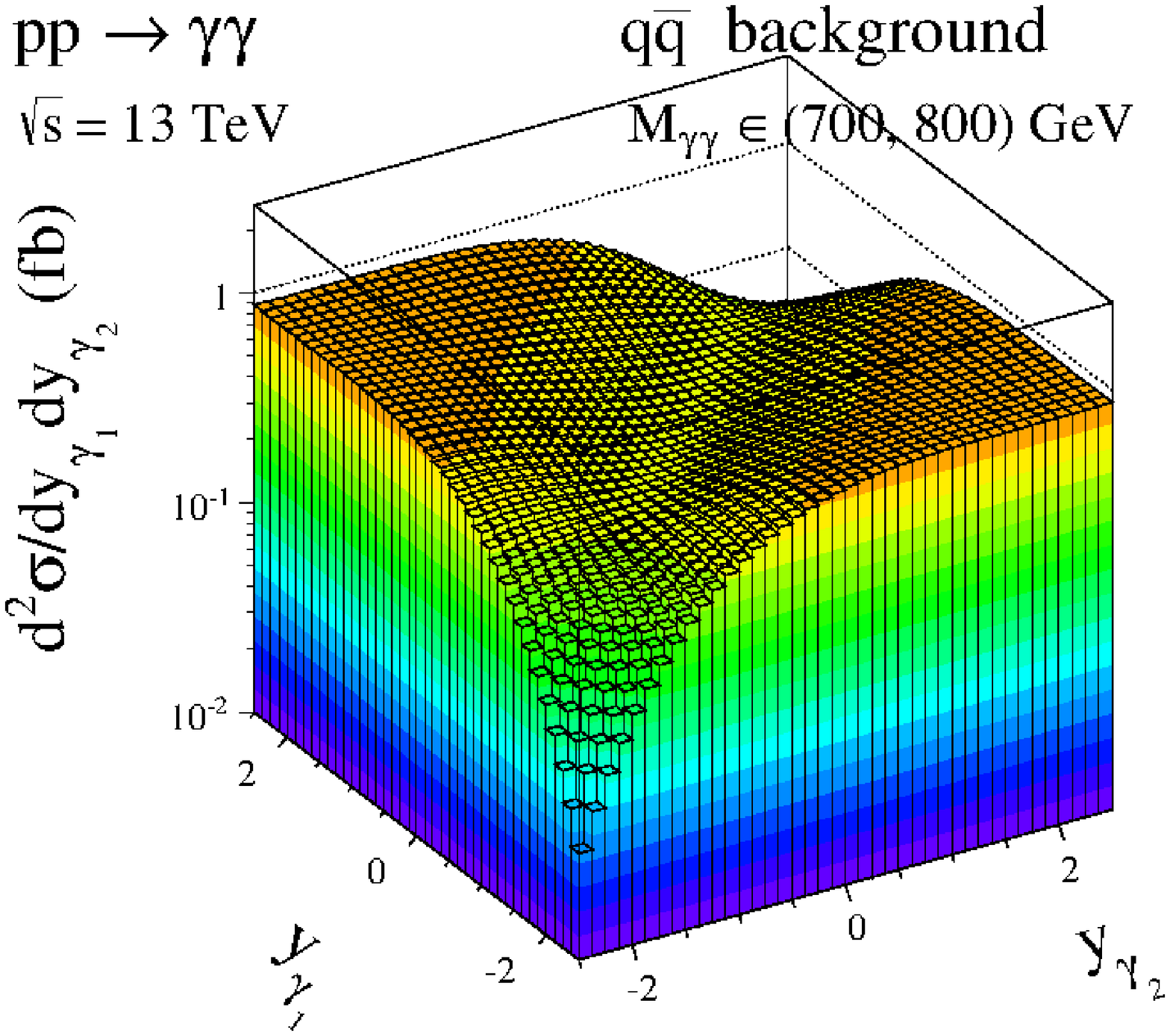}
\includegraphics[width=0.3\textwidth]{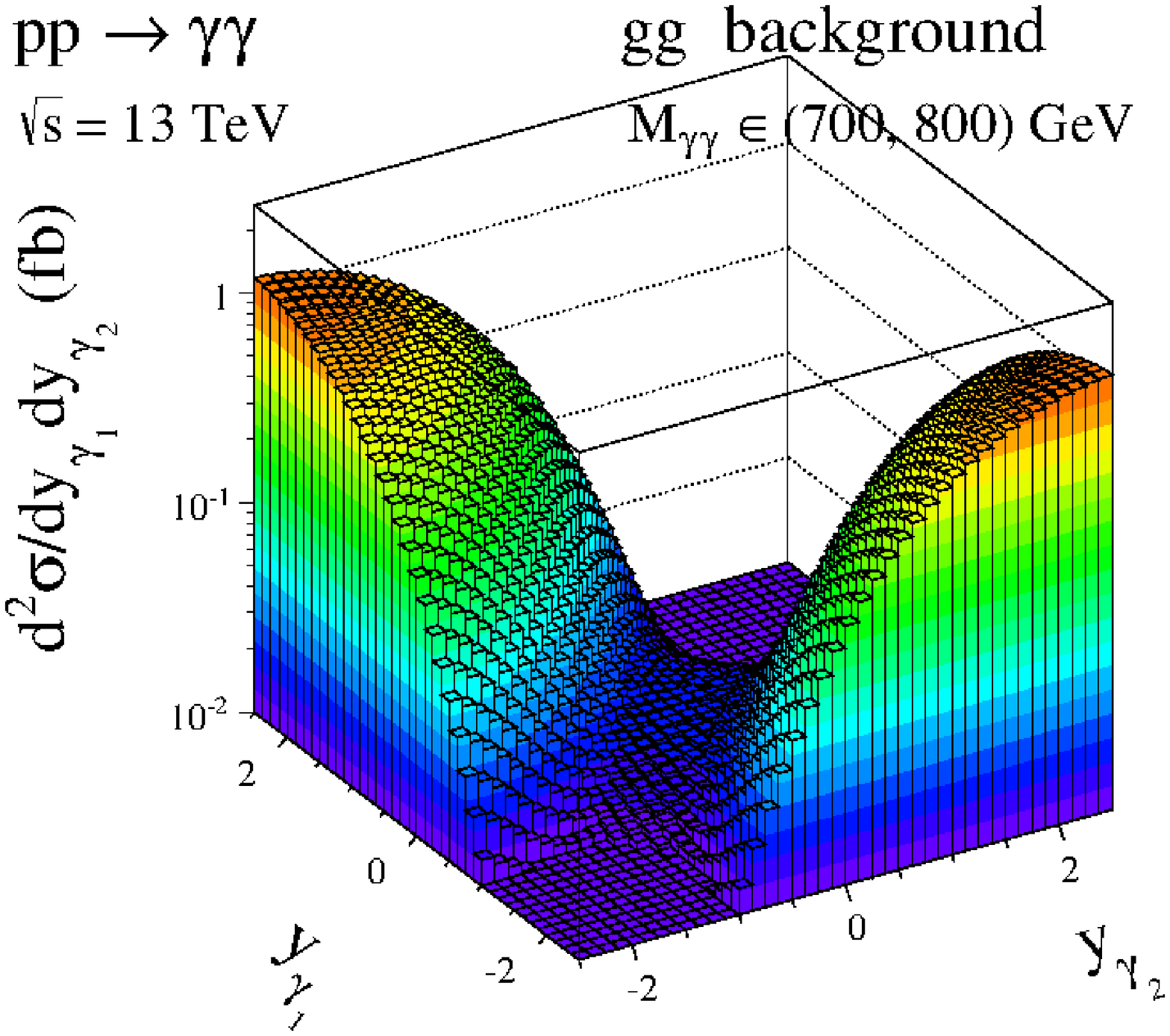}\\
\includegraphics[width=0.3\textwidth]{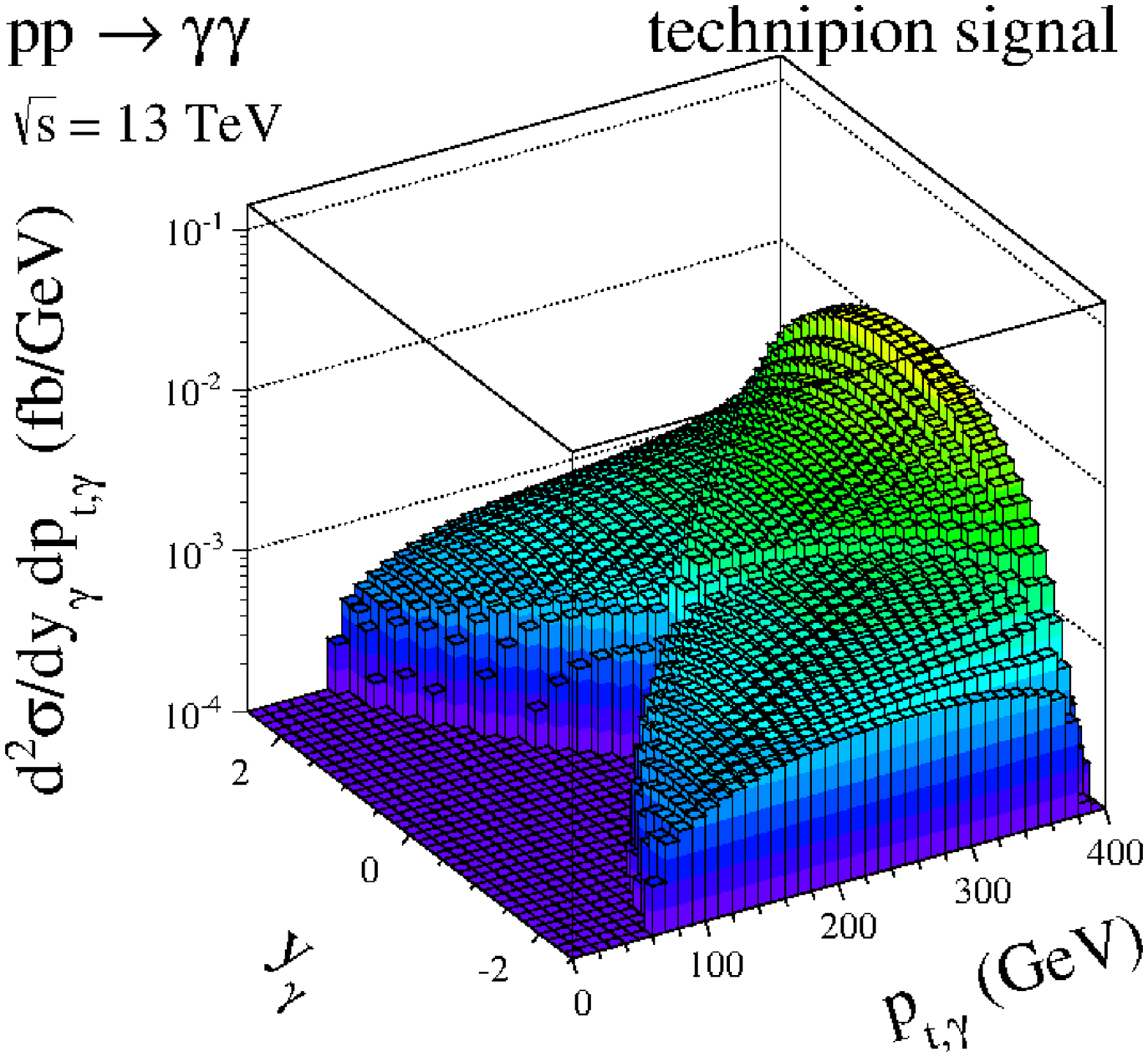}
\includegraphics[width=0.3\textwidth]{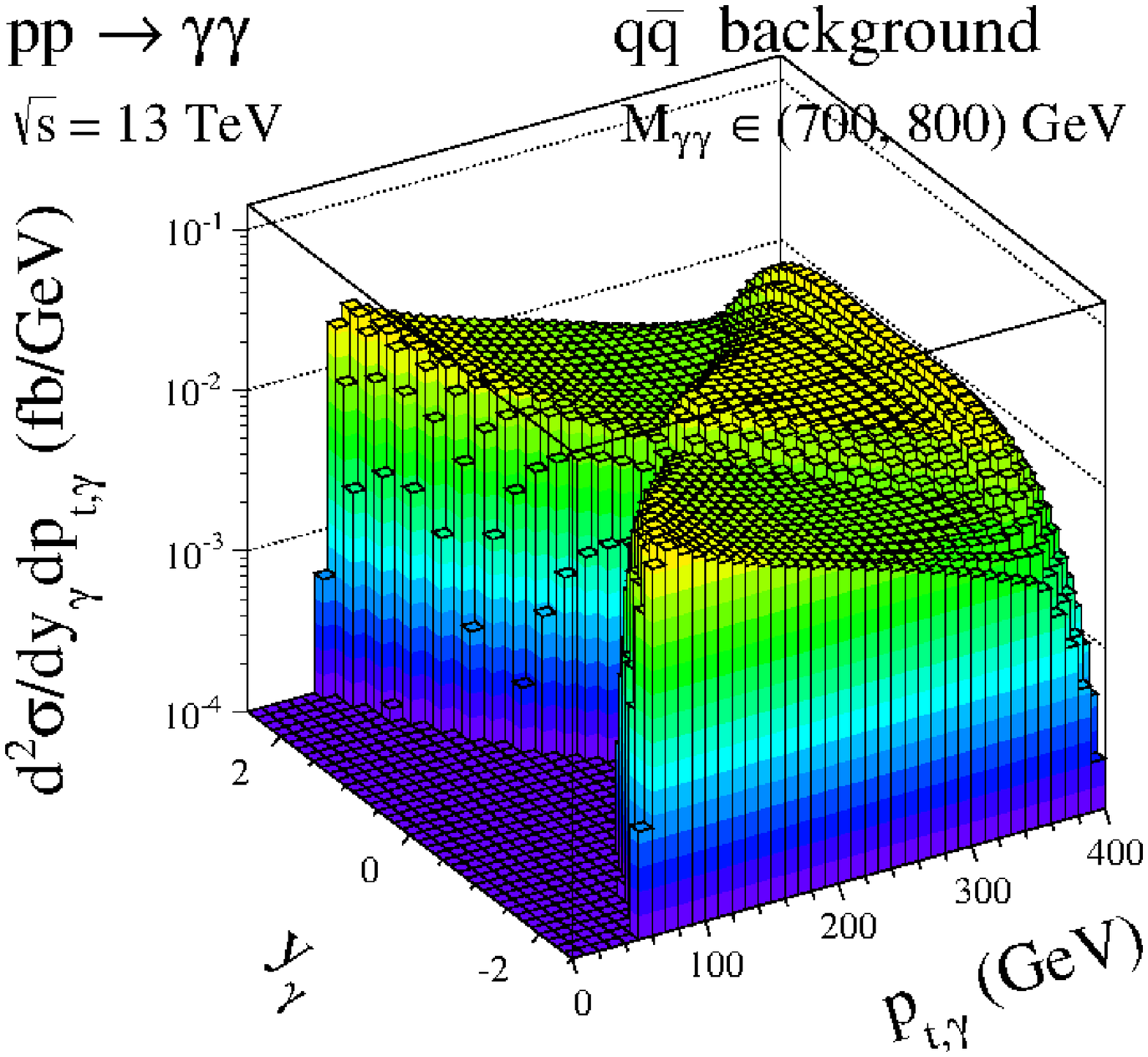}
\includegraphics[width=0.3\textwidth]{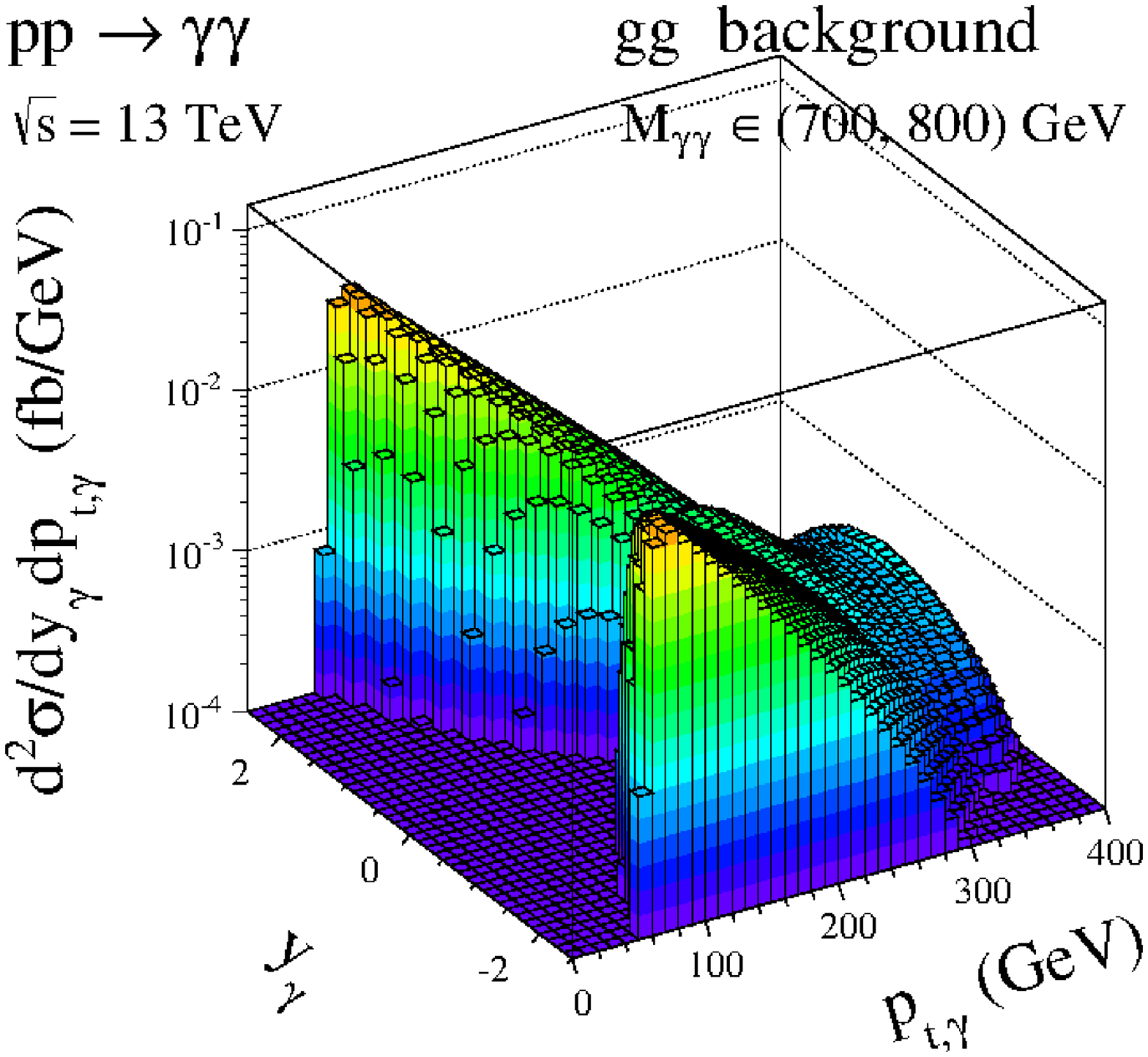}
  \caption{\label{fig:y3y4}
  \small
Two-dimensional distributions in rapidity of photons (top panels)
and distributions in photon rapidity 
and transverse momentum (bottom panels).
The signal (technipion) contribution 
with elastic and inelastic photon fluxes
and the background contributions 
in the diphoton invariant mass range 
$M_{\gamma \gamma} \in (700,800)$~GeV are shown.
In the technipion calculation $g_{TC}$ = 20 was used. 
}
\end{figure}

\begin{figure}[!ht]
\includegraphics[width=0.4\textwidth]{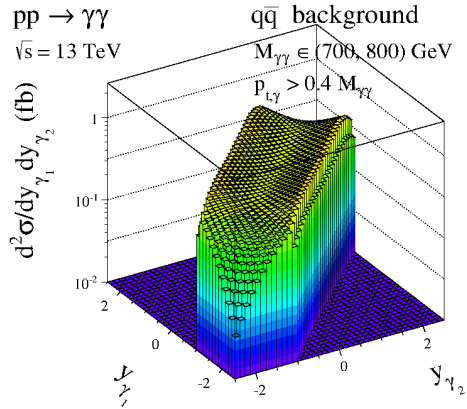}
  \caption{\label{fig:y3y4_ptcuts}
  \small
Two-dimensional distribution
for the $q\bar{q}$-annihilation contribution at $\sqrt{s}$ = 13 TeV
in the diphoton invariant mass range $M_{\gamma \gamma} \in (700,800)$~GeV
with extra limitations on both photon transverse momenta
$p_{t,\gamma} > 0.4\,M_{\gamma \gamma}$.
}
\end{figure}

Let us consider now two-dimensional distributions for the technipion signal 
and the $q\bar{q}$ and $gg$ background contributions at $\sqrt{s}$ = 13 TeV.
In Fig.~\ref{fig:y3y4} we show 
the distributions in photon rapidities limiting to $|y_{\gamma}|<2.5$
and in the rapidity and transverse momentum of one of outgoing photons.
Our signal obtained in the VTC model strongly
contributes at midrapidities $y_{\gamma} \approx 0$
while the background contributions have maximum 
in the regions $(y_{\gamma_{1}},y_{\gamma_{2}}) \approx (\pm 2.5,\mp 2.5)$.
The $q\bar{q}$ component of background
has on average larger transverse momenta of photons than the $gg$ component.
In Fig.~\ref{fig:y3y4_ptcuts} we show
the dominant $q\bar{q}$ background contribution
with extra limitations on both photon transverse momenta
$p_{t,\gamma} > 0.4\,M_{\gamma \gamma}$
that are inspired by the recent ATLAS analysis at $\sqrt{s}$ = 13 TeV \cite{ATLAS:2015}.
The experimental cuts $p_{t,\gamma} > 0.4\,M_{\gamma \gamma}$
significantly decrease the $gg$ and $q\bar{q}$ background cross sections
(in the region $M_{\gamma \gamma} \in (700,800)$~GeV)
from $\sigma_{gg} = 4.95$~fb to 0.17~fb
and from $\sigma_{q\bar{q}} = 15.05$~fb to 5.22~fb, respectively,
and lead to a rather small damping for the signal contribution
from $\sigma_{\tilde{\pi}^{0}} = 3.91$ to 2.62~fb.

\begin{figure}[!ht]
\includegraphics[width=0.45\textwidth]{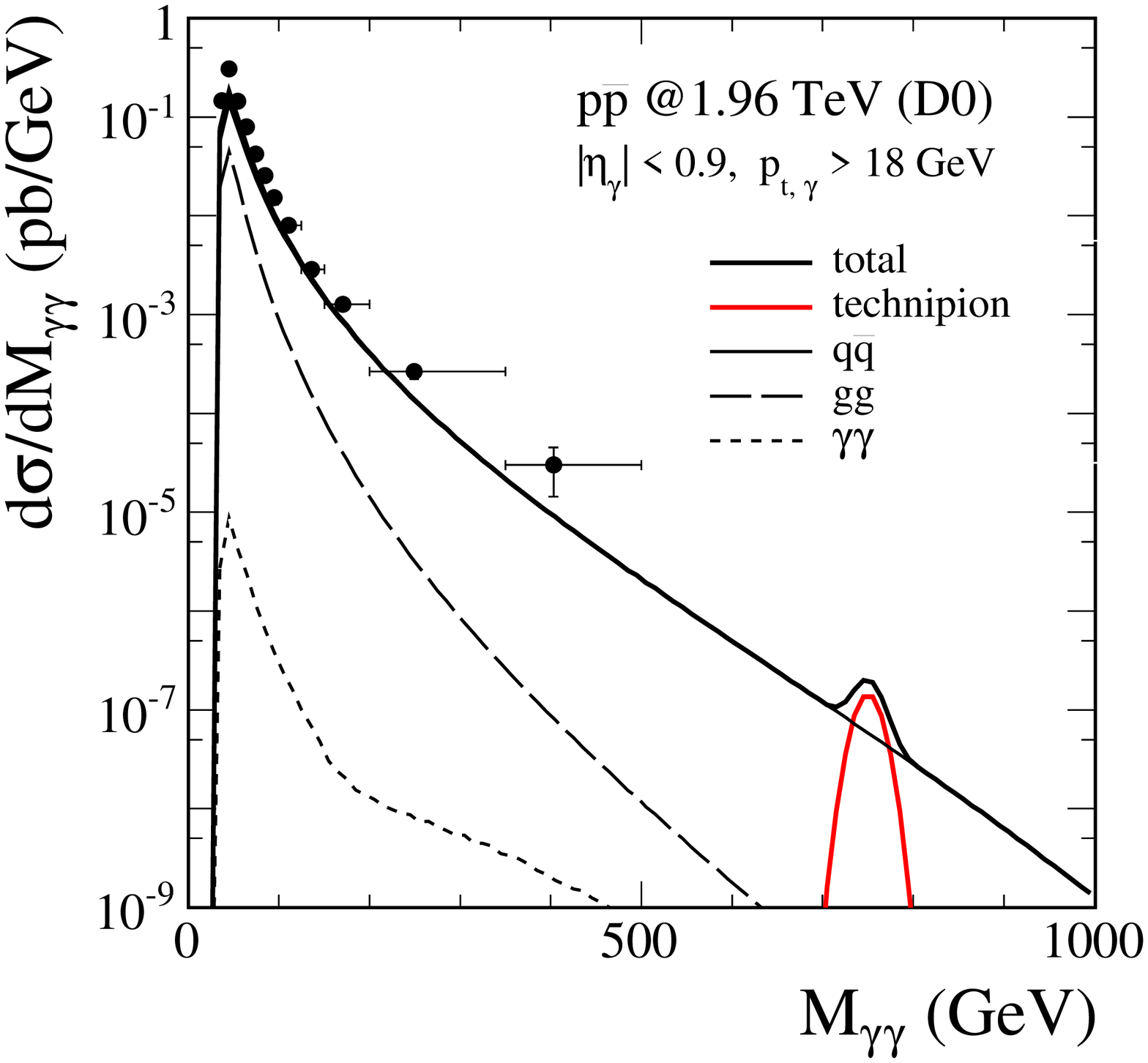}
\includegraphics[width=0.45\textwidth]{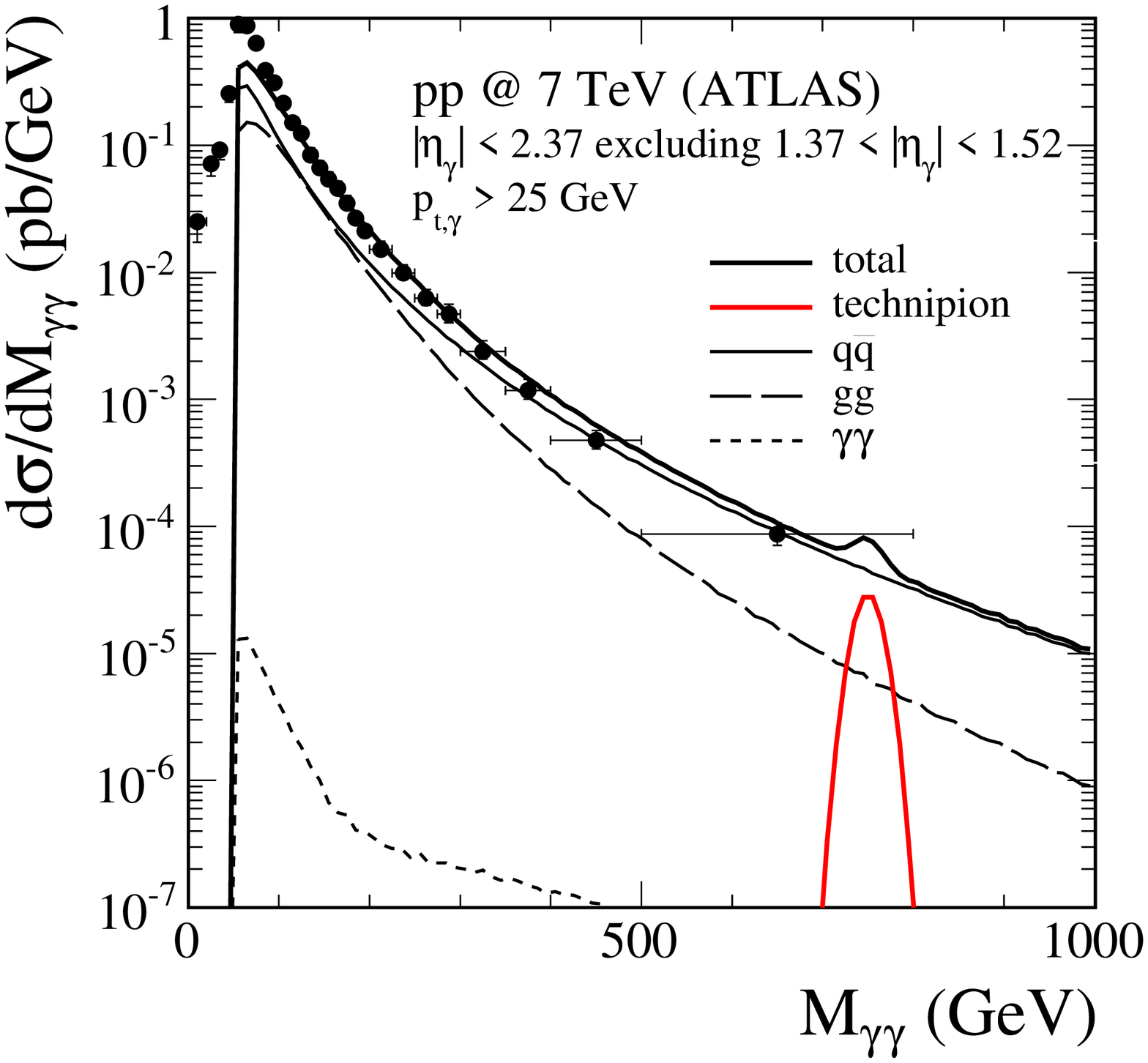}
\includegraphics[width=0.45\textwidth]{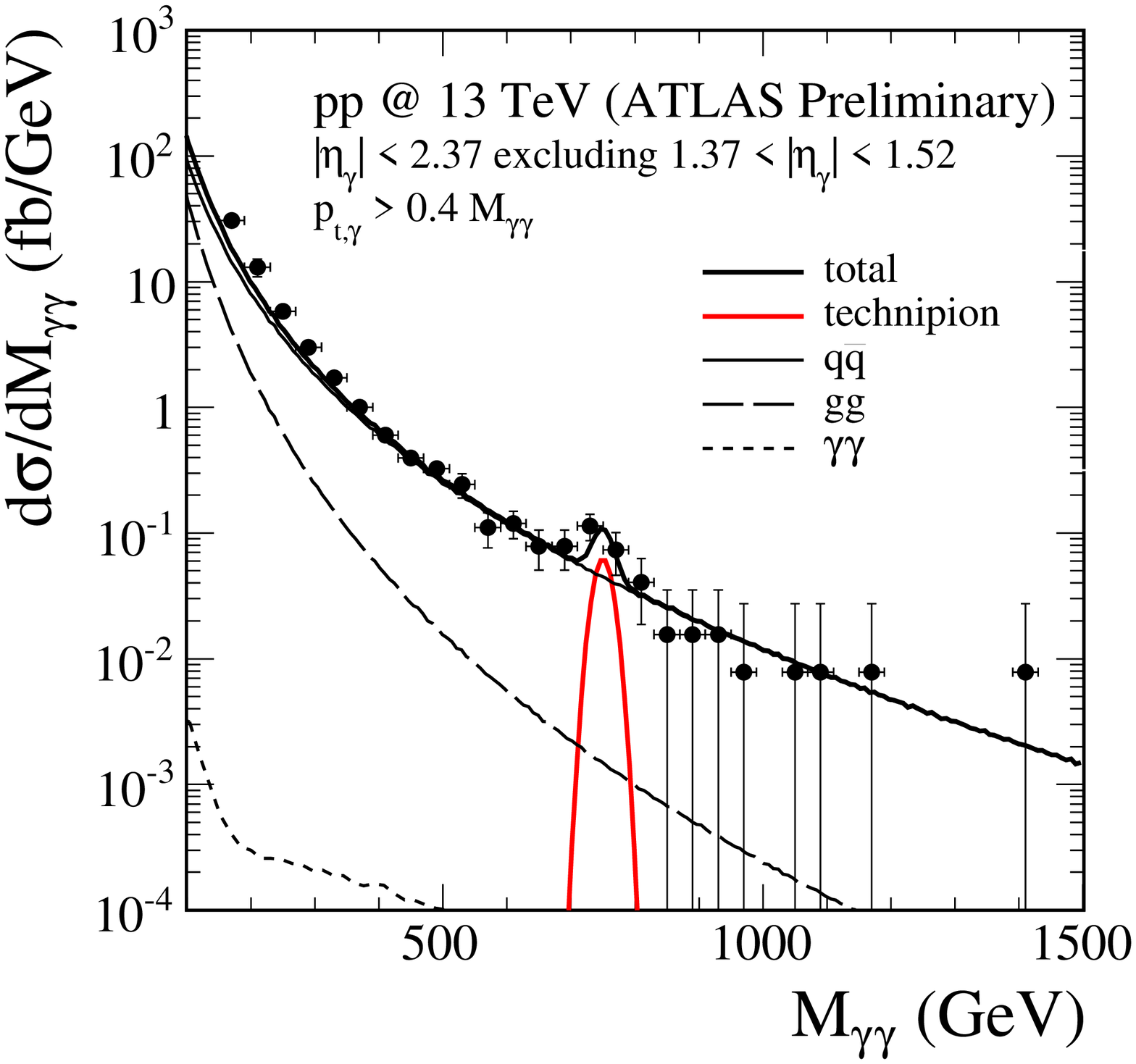}
\includegraphics[width=0.45\textwidth]{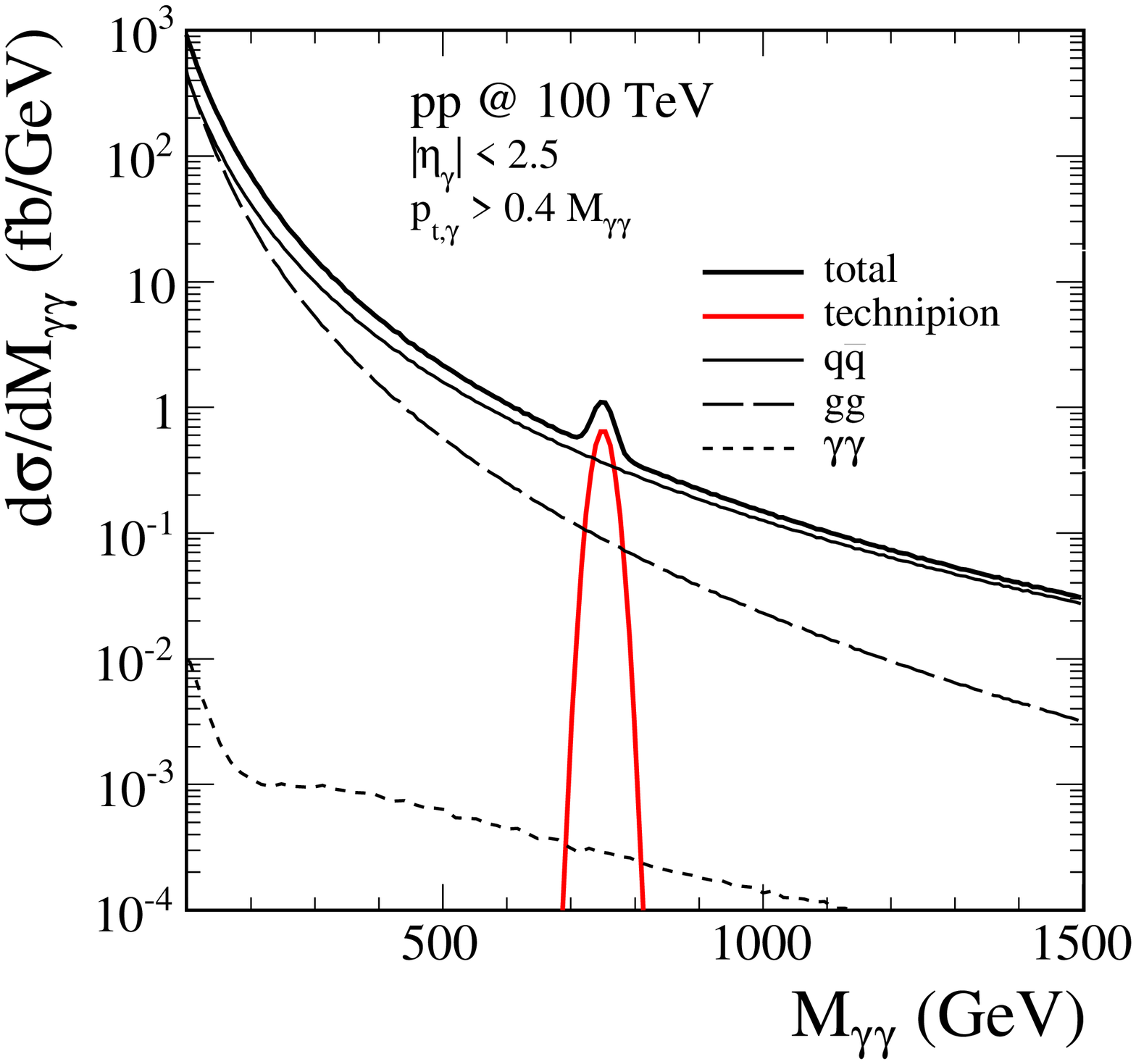}
  \caption{\label{fig:M_gamgam}
  \small
The two-photon invariant mass distributions 
for different background contributions
and the signal-technipion predictions obtained in the VTC model
including experimental cuts.
For comparison, the experimental data from D0 \cite{Abazov:2013pua} at $\sqrt{s}$ = 1.96 TeV,
ATLAS at $\sqrt{s}$ = 7 TeV \cite{Aad:2012tba}, 
the recent ATLAS data at $\sqrt{s}$ = 13 TeV \cite{ATLAS:2015}
and our prediction for Future Circular Collider are presented.
}
\end{figure}

The most important is the distribution in diphoton invariant mass
where the signal was observed by the ATLAS and CMS Collaborations. 
In Fig.~\ref{fig:M_gamgam} we show four examples relevant 
for different experiments using their kinematic conditions:
D0 at $\sqrt{s}$ = 1.96 TeV \cite{Abazov:2013pua},
ATLAS at $\sqrt{s}$ = 7 TeV \cite{Aad:2012tba} (see also CMS data in \cite{Chatrchyan:2014fsa}),
and at $\sqrt{s}$ = 13 TeV \cite{ATLAS:2015}.
We show both signal and background (see the previous section) contributions.
Clearly the $q \bar q$ annihilation contribution dominates, especially
at large invariant masses in the surrounding of the signal. 
In the photon-induced contributions all components 
(elastic-elastic, elastic-inelastic, inelastic-inelastic) were taken into account.
The $\gamma \gamma$ contribution shows two slopes. 
For $M_{\gamma \gamma} >$ 200 GeV the boxes with $W$ boson dominate \cite{Lebiedowicz:2013fta}.
The experimental smearing effect leads to a significant modification of the sharp peaks, 
e.g. for the Higgs signal, see Ref.~\cite{Maciula:2010tv}.
The experimental invariant mass resolution was included for the signal-technipion calculations
in the following simple way
\begin{eqnarray}
\frac{d \sigma}{dM_{\gamma\gamma}} 
= \sigma_{\tilde{\pi}^{0}} \, \frac{1}{\sqrt{2 \pi} \sigma} \exp\left({\frac{-(M_{\gamma\gamma} - m_{\tilde{\pi}^{0}})^{2}}{2 \sigma^{2}}}\right)\,.
\label{resolution}
\end{eqnarray}
In the calculation we take $\sigma = 15$~GeV assuming $\sigma/m_{\tilde{\pi}^{0}} \sim$ 2\%.
In Eq.~(\ref{resolution}) we take $\sigma_{\tilde{\pi}^{0}}$ = 0.005~fb, 1.09~fb, 2.36~fb, 24.83~fb
corresponding to $\sqrt{s}$ = 1.96, 7, 13, 100 TeV, respectively,
including the relevant kinematical cuts shown in the panels of Fig.~\ref{fig:M_gamgam}.
The values of cross sections above were obtained from Eq.~(\ref{inclusive_gamgam_signal})and $g_{TC} = 20$.

\subsection{Comparison of WTC signal with the existing data for dijets production}

\begin{figure}[!ht]
\includegraphics[width=0.45\textwidth]{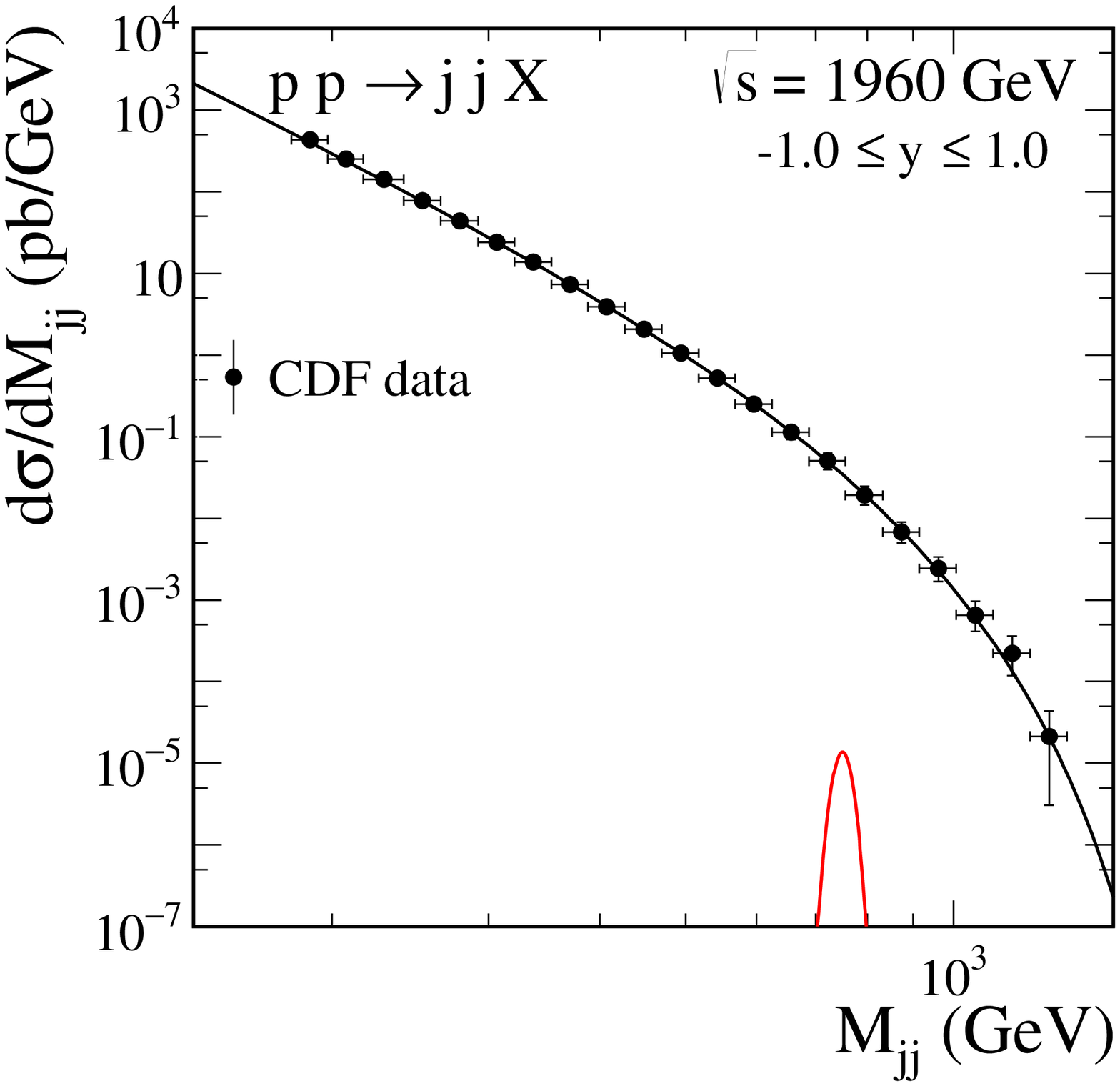}
\includegraphics[width=0.45\textwidth]{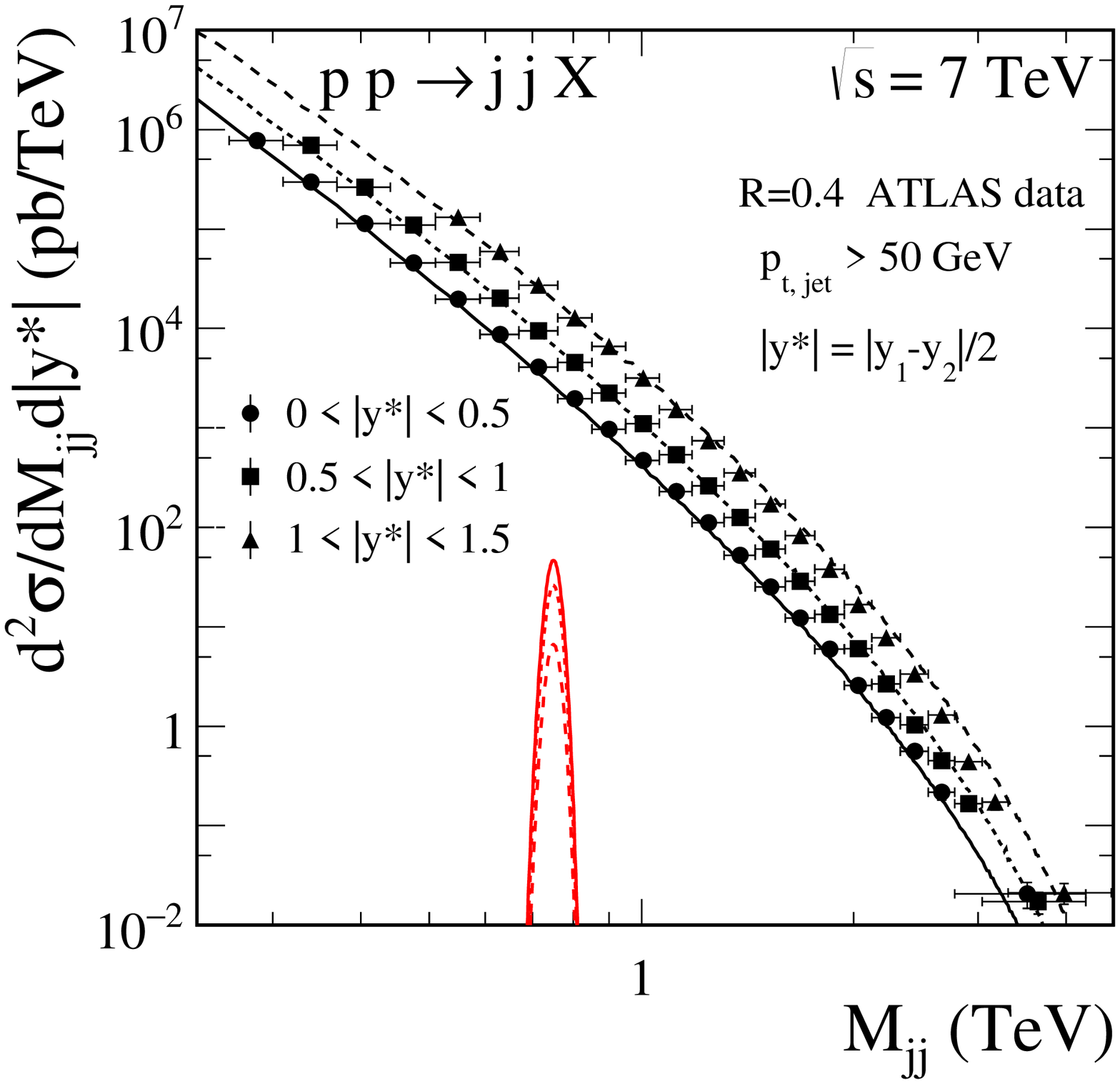}
  \caption{\label{fig:M_jj}
  \small
Dijet invariant mass distribution for the one-family walking technipion (red lines). 
We show results of both the CDF \cite{Aaltonen:2008dn} (left panel) and 
ATLAS \cite{Aad:2013tea} (right panel) Collaborations. 
Please note different order of lines for the signal and the background contributions.
In this calculation we use leading-order MSTW08 PDFs \cite{Martin:2009iq}
and $\mu_{F}^{2} = p_{t,jet}^{2}$.
}
\end{figure}

Finally in Fig.~\ref{fig:M_jj} we show the one-family WTC signal of technipion 
in the dijet final state together with the 
CDF \cite{Aaltonen:2008dn} and ATLAS \cite{Aad:2013tea} data.
For corresponding diagram of production mechanism see the right panel of Fig.~\ref{fig:diagrams_gammagamma_fstate}.
In the case of the ATLAS data we show separate results for
different ranges of an auxiliary variable: 
$y^* = |y_1 - y_2|/2$.
In both cases the translated signal is below the experimental data. 
This means that the model cannot be excluded. 
Although when the model total (gluon-gluon) decay width of 1.2 GeV 
(see Eq.~(\ref{partial_decay_widths_gg})), 
was used and experimental
resolution was ignored some tension could be probably observed.
In an older version of the model \cite{Jia:2012kd} (top quark mass solely generated
through the ETC (extended technicolor approach)) also a strong coupling
of isoscalar technipion to top quarks was considered, which would lead
to a (too)strong signal in the $t {\overline t}$ channel. 
However, the top quark mass may arise also from other mechanisms like top condensation for instance. 
In our analysis here we followed therefore the recent version 
of the model \cite{Matsuzaki:2015che}, where the coupling to top quarks is totally neglected.

We show also standard dijet background contribution calculated in
leading-order pQCD, which is sufficient for the present precision
of searches for the signal of new physics.
To improve the precision of the description of the data a $K$-factor
simulating higher-order corrections was applied in the case of the CDF
data.  
In all cases the estimated signal including, similarly as for the
diphoton final state, a 2\% dijet invariant mass resolution, 
is substantially below experimental data and standard dijet contribution.
We observe a clear tendency that the signal-to-background ratio improves (increases)
when going to small values of $y^*$ (please note different ordering of
lines for signal and background contributions).

\section{Conclusions and Outlook}
\label{sec:Conclusions}

In the present paper, we have discussed a possibility that recently
observed by the ATLAS and CMS Collaborations diphoton signal 
at invariant mass $M_{\gamma \gamma} \approx$ 750 GeV is a technipion. 
The main emphasis was put on chirally-symmetric (vector-like) technicolor (VTC) model with 
two mass degenerate (techni)flavours. In this model only $\gamma \gamma$, $\gamma Z$ and $Z Z$ 
couplings are possible. Therefore the decay width is rather small 
$\Gamma_{tot} \ll$ 1 GeV, unless some other decays into stable objects
-- candidate(s) for dark matter, are considered.

We have discussed in detail the production mechanisms within the VTC model. In the present
analysis we have included only photon initiated processes which should be sufficient to estimate parameters 
of the model. In some modern parton models also photons are included as partons
in the proton. In this model there is a rich pattern of electroweak
contributions. We have considered $2 \to 1$, $2 \to 2$ and $2 \to 3$ type
of subprocesses and discussed also some interesting technical details of the calculation.
We have found that they give similar contributions
to the hadronic cross section. In order to describe the observed ``signal''
we had to adjust model coupling of techniquarks to the neutral technipion. 
Including the photon initiated processes we have found
that $10 < g_{TC} < 20$ in order to describe experimentally obtained
cross section.
Several differential distributions have been presented for 
the $\gamma \gamma$ induced processes.
Some specific features of the $2 \to 3$ calculations have been
illustrated and discussed.

Having adjusted the $g_{TC}$ parameter to reproduce the observed signal 
(see Refs.~\cite{ATLAS:2015,CMS:2015dxe}),
we have made predictions for the Tevatron, Run-I LHC 
and for the Future Circular Collider. The predictions for the Tevatron and
ATLAS (at $\sqrt{s}$ = 7 TeV) have been discussed in the context 
of existing data in the diphoton channel.
We have concluded that the cross section for energies lower than 13 TeV
are so small (below background for integrated luminosity limit) 
that the signal could not be observed. 

We have shown that in order to improve the signal-to-background ratio
one could try to measure the hypothetical technipion signal together
with one or two-jets. With mild cuts (small lower cuts on jet transverse
momenta) the cross section is reduced only by not more than one order 
of magnitude. The reduction is of course larger than in the case when 
we require the presence of two jets. Already the requirement
of only one extra jet looks promising. This issue requires further 
detailed studies, which could be done in the case when the ``signal'' 
is confirmed at the LHC with a better statistics.

Fixing the relevant model coupling constant we have also made 
predictions for purely exclusive case, called here the
elastic-elastic case for the sake of brevity.
We have predicted that the corresponding cross section should be
of the order of 0.2~fb at $\sqrt{s} = 13$~TeV.
To focus on such a case one has to measure technipion (two photons) 
in the central detectors as well as both protons in forward directions. 
Relevant ``forward detectors'' are being installed both by the
ATLAS and CMS Collaborations.  Unfortunately, our predicted cross section
seems too small to allow for interesting studies of spin and parity
of the resonance discussed very recently \cite{Harland-Lang:2016qjy}.

For comparison we have considered also an alternative one-family walking technicolor (WTC) model. 
In this model gluon-gluon fusion is the dominant production
mechanism of the assumed isoscalar technipion. Such an object decays also
to the two-gluon (dijet) final state.
We have presented predictions of this model for the dijet final state 
and found corresponding signal significantly below the CDF and ATLAS data
as well as below the standard dijet background.
We have found that signal-to-background ratio strongly depends
on so-called $y^*$ variable. The smaller $y^{*}$ the larger the
signal-to-background ratio is.
The WTC model could be further verified in the future by
considering a four-jet analysis $p p \to ({\tilde \pi}^0 \to j j) j j$
in a similar way as have been done for the VTC model for 
$p p \to ({\tilde \pi}^0 \to \gamma \gamma) jj$. 
The background would be then the QCD production of four jets 
(two forward and two central) which is now calculable for leading and
next-to-leading collinear approximation (see, e.g.~\cite{Bern:2011ep})
as well as in the $k_t$-factorization approach \cite{Kutak:2016mik}.
These clearly goes beyond the scope of the present paper
and will be done elsewhere.
In the WTC model the two-jet signal with two extra jets
would be reduced by an order of magnitude while the background by
at least two orders of magnitude.

In summary, neither of the considered technicolor models can be excluded
by the present world $\gamma \gamma$ and dijet experimental data.
We have shown that the diphoton signal in the VTC model
is consistent with all existing diphoton experimental data.
The experimental signal for $Z \gamma$ and $ZZ$
(similar production rate as for $\gamma\gamma$ is predicted in the VTC model)
is much smaller as it includes branching fraction(s) for the decay of $Z$ boson(s)
into e.g. leptons. A combined analysis of different final states
(leptons, jets) is required but is of course much more complicated.
The WTC model predicts the existence of many states such as
isotriplet of states called by the authors $P^{0}$, $P^{+}$, $P^{-}$ \cite{Kurachi:2014xla}.
They predict the mass of the state(s) at $M \thickapprox 900$~GeV.
The neutral member of the triplet could also decay
into the $\gamma \gamma$ final state.
Due to its quantum numbers it cannot be, however,
produced by the gluon-gluon fusion so the corresponding
signal would be much weaker than for the isoscalar state 
discussed in the present paper and therefore difficult to be observed.
The isotriplet technirho meson was suggested \cite{Fukano:2015hga}
as a possible explanation of the diboson enhancement observed
by the ATLAS Collaboration \cite{Aad:2015owa,Aad:2015ipg} at $M \thickapprox 2$~TeV.

In the present paper we have made a consistency analysis of two selected
technicolor models as far as technipion production is considered.
A similar, rather straightforward, analysis can be made also for axion
production. For example for the model considered in Ref.~\cite{Barrie:2016ntq},
the axion is produced also via photon-photon fusion and the methods
discussed here apply.

\acknowledgments
We are indebted to Wolfgang Sch{\"a}fer for a discussion
on $\gamma \gamma$ induced processes and to Shinya Matsuzaki for
explanation of several details of their works on walking technicolor model.
This research was partially supported by the Polish National Science
Centre Grant No. DEC-2014/15/B/ST2/02528 (OPUS) and by the Centre 
for Innovation and Transfer of Natural Sciences and Engineering Knowledge in Rzesz{\'o}w. 
R. P. was partially supported by the Swedish Research Council Grant No. 2013-4287.

\bibliography{refs}

\providecommand{\href}[2]{#2}\begingroup\raggedright\begin{thebibliography}{10}

\bibitem{ATLAS:2015}
G.~Aad {\em et~al.}, (ATLAS Collaboration), {\em {Search for resonances
  decaying to photon pairs in 3.2 fb$^{-1}$ of $pp$ collisions at $\sqrt{s} =
  13$ TeV with the ATLAS detector},} ATLAS-CONF-2015-081,
ATLAS-CONF-2015-081 (2015) .

\bibitem{CMS:2015dxe}
V.~Khachatryan {\em et~al.}, (CMS Collaboration), {\em {Search for new physics
  in high mass diphoton events in proton-proton collisions at 13 TeV},}
  CMS-PAS-EXO-15-004,
CMS PAS EXO-15-004 (2015) .

\bibitem{Aad:2014ioa}
G.~Aad {\em et~al.}, (ATLAS Collaboration), {\em {Search for Scalar Diphoton
  Resonances in the Mass Range $65-600$ GeV with the ATLAS Detector in $pp$
  Collision Data at $\sqrt{s}$ = 8 $TeV$},}
  \href{http://dx.doi.org/10.1103/PhysRevLett.113.171801}{Phys. Rev. Lett.
  {\bfseries 113} no.~17, (2014) 171801},
\href{http://arxiv.org/abs/1407.6583}{{arXiv:1407.6583 [hep-ex]}}.

\bibitem{Ellis:2015oso}
J.~Ellis, S.~A.~R. Ellis, J.~Quevillon, V.~Sanz, and T.~You, {\em {On the
  Interpretation of a Possible $\sim 750$ GeV Particle Decaying into $\gamma
  \gamma$},}
\href{http://arxiv.org/abs/1512.05327}{{arXiv:1512.05327 [hep-ph]}}.

\bibitem{Franceschini:2015kwy}
R.~Franceschini, G.~F. Giudice, J.~F. Kamenik, M.~McCullough, A.~Pomarol,
  R.~Rattazzi, M.~Redi, F.~Riva, A.~Strumia, and R.~Torre, {\em {What is the
  gamma gamma resonance at 750 GeV?},}
\href{http://arxiv.org/abs/1512.04933}{{arXiv:1512.04933 [hep-ph]}}.

\bibitem{Csaki:2015vek}
C.~Cs{\'a}ki, J.~Hubisz, and J.~Terning, {\em {Minimal model of a diphoton
  resonance: Production without gluon couplings},}
  \href{http://dx.doi.org/10.1103/PhysRevD.93.035002}{Phys. Rev. {\bfseries
  D93} no.~3, (2016) 035002},
\href{http://arxiv.org/abs/1512.05776}{{arXiv:1512.05776 [hep-ph]}}.

\bibitem{Fichet:2015vvy}
S.~Fichet, G.~von Gersdorff, and C.~Royon, {\em {Scattering Light by Light at
  750 GeV at the LHC},}
\href{http://arxiv.org/abs/1512.05751}{{arXiv:1512.05751 [hep-ph]}}.

\bibitem{Fichet:2016pvq}
S.~Fichet, G.~von Gersdorff, and C.~Royon, {\em {Measuring the diphoton
  coupling of a 750 GeV resonance},}
\href{http://arxiv.org/abs/1601.01712}{{arXiv:1601.01712 [hep-ph]}}.

\bibitem{Luszczak:2011uh}
M.~{\L}uszczak, R.~Maciu{\l}a, and A.~Szczurek, {\em {Subdominant terms in the
  production of $c \bar{c}$ pairs in proton-proton collisions},}
  \href{http://dx.doi.org/10.1103/PhysRevD.84.114018}{Phys. Rev. {\bfseries
  D84} (2011) 114018},
\href{http://arxiv.org/abs/1109.5930}{{arXiv:1109.5930 [hep-ph]}}.

\bibitem{Maciula:2010vc}
R.~Maciu{\l}a, R.~Pasechnik, and A.~Szczurek, {\em {Exclusive $b \bar b$ pair
  production and irreducible background to the exclusive Higgs boson
  production},} \href{http://dx.doi.org/10.1103/PhysRevD.82.114011}{Phys. Rev.
  {\bfseries D82} (2010) 114011},
\href{http://arxiv.org/abs/1006.3007}{{arXiv:1006.3007 [hep-ph]}}.

\bibitem{Maciula:2010tv}
R.~Maciu{\l}a, R.~Pasechnik, and A.~Szczurek, {\em {Central exclusive
  quark-antiquark dijet and Standard Model Higgs boson production in
  proton-(anti)proton collisions},}
  \href{http://dx.doi.org/10.1103/PhysRevD.83.114034}{Phys. Rev. {\bfseries
  D83} (2011) 114034},
\href{http://arxiv.org/abs/1011.5842}{{arXiv:1011.5842 [hep-ph]}}.

\bibitem{daSilveira:2014jla}
G.~G. da~Silveira, L.~Forthomme, K.~Piotrzkowski, W.~Sch{\"a}fer, and
  A.~Szczurek, {\em {Central $\mu^{+}$ $\mu^{-}$ production via photon-photon
  fusion in proton-proton collisions with proton dissociation},}
  \href{http://dx.doi.org/10.1007/JHEP02(2015)159}{JHEP {\bfseries 02} (2015)
  159},
\href{http://arxiv.org/abs/1409.1541}{{arXiv:1409.1541 [hep-ph]}}.

\bibitem{Luszczak:2015aoa}
M.~{\L}uszczak, W.~Sch{\"a}fer, and A.~Szczurek, {\em {Two-photon dilepton
  production in proton-proton collisions: two alternative approaches},}
\href{http://arxiv.org/abs/1510.00294}{{arXiv:1510.00294 [hep-ph]}}.

\bibitem{Lebiedowicz:2012gg}
P.~Lebiedowicz, R.~Pasechnik, and A.~Szczurek, {\em {QCD diffractive mechanism
  of exclusive $W^+W^-$ pair production at high energies},}
  \href{http://dx.doi.org/10.1016/j.nuclphysb.2012.09.014}{Nucl. Phys.
  {\bfseries B867} (2013) 61--81},
\href{http://arxiv.org/abs/1203.1832}{{arXiv:1203.1832 [hep-ph]}}.

\bibitem{Luszczak:2014mta}
M.~{\L}uszczak, A.~Szczurek, and C.~Royon, {\em {$W^+ W^-$ pair production in
  proton-proton collisions: small missing terms},}
  \href{http://dx.doi.org/10.1007/JHEP02(2015)098}{JHEP {\bfseries 02} (2015)
  098},
\href{http://arxiv.org/abs/1409.1803}{{arXiv:1409.1803 [hep-ph]}}.

\bibitem{Lebiedowicz:2015cea}
P.~Lebiedowicz and A.~Szczurek, {\em {Exclusive production of heavy charged
  Higgs boson pairs in the $p p \to p p H^+ H^-$ reaction at the LHC and a
  future circular collider},}
  \href{http://dx.doi.org/10.1103/PhysRevD.91.095008}{Phys. Rev. {\bfseries
  D91} (2015) 095008},
\href{http://arxiv.org/abs/1502.03323}{{arXiv:1502.03323 [hep-ph]}}.

\bibitem{Weinberg:1975gm}
S.~Weinberg, {\em {Implications of Dynamical Symmetry Breaking},}
\href{http://dx.doi.org/10.1103/PhysRevD.13.974}{Phys. Rev. {\bfseries D13}
  (1976) 974--996}.

\bibitem{Susskind:1978ms}
L.~Susskind, {\em {Dynamics of Spontaneous Symmetry Breaking in the
  Weinberg-Salam Theory},}
\href{http://dx.doi.org/10.1103/PhysRevD.20.2619}{Phys. Rev. {\bfseries D20}
  (1979) 2619--2625}.

\bibitem{Eichten:1979ah}
E.~Eichten and K.~D. Lane, {\em {Dynamical Breaking of Weak Interaction
  Symmetries},}
\href{http://dx.doi.org/10.1016/0370-2693(80)90065-9}{Phys. Lett. {\bfseries
  B90} (1980) 125--130}.

\bibitem{Peskin:1990zt}
M.~E. Peskin and T.~Takeuchi, {\em {A New constraint on a strongly interacting
  Higgs sector},}
\href{http://dx.doi.org/10.1103/PhysRevLett.65.964}{Phys. Rev. Lett. {\bfseries
  65} (1990) 964--967}.

\bibitem{Peskin:1991sw}
M.~E. Peskin and T.~Takeuchi, {\em {Estimation of oblique electroweak
  corrections},}
\href{http://dx.doi.org/10.1103/PhysRevD.46.381}{Phys. Rev. {\bfseries D46}
  (1992) 381--409}.

\bibitem{Galloway:2010bp}
J.~Galloway, J.~A. Evans, M.~A. Luty, and R.~A. Tacchi, {\em {Minimal Conformal
  Technicolor and Precision Electroweak Tests},}
  \href{http://dx.doi.org/10.1007/JHEP10(2010)086}{JHEP {\bfseries 10} (2010)
  086},
\href{http://arxiv.org/abs/1001.1361}{{arXiv:1001.1361 [hep-ph]}}.

\bibitem{Arbey:2015exa}
A.~Arbey, G.~Cacciapaglia, H.~Cai, A.~Deandrea, S.~Le~Corre, and F.~Sannino,
  {\em {Fundamental Composite Electroweak Dynamics: Status at the LHC},}
\href{http://arxiv.org/abs/1502.04718}{{arXiv:1502.04718 [hep-ph]}}.

\bibitem{Hill:2002ap}
C.~T. Hill and E.~H. Simmons, {\em {Strong dynamics and electroweak symmetry
  breaking},} \href{http://dx.doi.org/10.1016/S0370-1573(03)00140-6}{Phys.
  Rept. {\bfseries 381} (2003) 235--402},
  \href{http://arxiv.org/abs/hep-ph/0203079}{{arXiv:hep-ph/0203079 [hep-ph]}}.
[Erratum: Phys. Rept.390,553(2004)].

\bibitem{Sannino:2009za}
F.~Sannino, {\em {Conformal Dynamics for TeV Physics and Cosmology},} Acta
  Phys. Polon. {\bfseries B40} (2009) 3533--3743,
\href{http://arxiv.org/abs/0911.0931}{{arXiv:0911.0931 [hep-ph]}}.

\bibitem{Kilic:2009mi}
C.~Kilic, T.~Okui, and R.~Sundrum, {\em {Vectorlike Confinement at the LHC},}
  \href{http://dx.doi.org/10.1007/JHEP02(2010)018}{JHEP {\bfseries 02} (2010)
  018},
\href{http://arxiv.org/abs/0906.0577}{{arXiv:0906.0577 [hep-ph]}}.

\bibitem{Pasechnik:2013bxa}
R.~Pasechnik, V.~Beylin, V.~Kuksa, and G.~Vereshkov, {\em {Chiral-symmetric
  technicolor with standard model Higgs boson},}
  \href{http://dx.doi.org/10.1103/PhysRevD.88.075009}{Phys. Rev. {\bfseries
  D88} no.~7, (2013) 075009},
\href{http://arxiv.org/abs/1304.2081}{{arXiv:1304.2081 [hep-ph]}}.

\bibitem{Lebiedowicz:2013fta}
P.~Lebiedowicz, R.~Pasechnik, and A.~Szczurek, {\em {Search for technipions in
  exclusive production of diphotons with large invariant masses at the LHC},}
  \href{http://dx.doi.org/10.1016/j.nuclphysb.2014.02.008}{Nucl. Phys.
  {\bfseries B881} (2014) 288--308},
\href{http://arxiv.org/abs/1309.7300}{{arXiv:1309.7300 [hep-ph]}}.

\bibitem{Pasechnik:2014ida}
R.~Pasechnik, V.~Beylin, V.~Kuksa, and G.~Vereshkov, {\em {Scalar technibaryon
  Dark Matter from vector-like SU(2) Technicolor},}
  \href{http://dx.doi.org/10.1142/S0217751X16500366}{Int.J.Mod.Phys. A
  {\bfseries 31} (2016) 1650036},
\href{http://arxiv.org/abs/1407.2392}{{arXiv:1407.2392 [hep-ph]}}.

\bibitem{Cacciapaglia:2014uja}
G.~Cacciapaglia and F.~Sannino, {\em {Fundamental Composite (Goldstone) Higgs
  Dynamics},} \href{http://dx.doi.org/10.1007/JHEP04(2014)111}{JHEP {\bfseries
  04} (2014) 111},
\href{http://arxiv.org/abs/1402.0233}{{arXiv:1402.0233 [hep-ph]}}.

\bibitem{Hietanen:2014xca}
A.~Hietanen, R.~Lewis, C.~Pica, and F.~Sannino, {\em {Fundamental Composite
  Higgs Dynamics on the Lattice: SU(2) with Two Flavors},}
  \href{http://dx.doi.org/10.1007/JHEP07(2014)116}{JHEP {\bfseries 07} (2014)
  116},
\href{http://arxiv.org/abs/1404.2794}{{arXiv:1404.2794 [hep-lat]}}.

\bibitem{Matsuzaki:2015che}
S.~Matsuzaki and K.~Yamawaki, {\em {750 GeV Diphoton Signal from One-Family
  Walking Technipion},}
\href{http://arxiv.org/abs/1512.05564}{{arXiv:1512.05564 [hep-ph]}}.

\bibitem{Molinaro:2015cwg}
E.~Molinaro, F.~Sannino, and N.~Vignaroli, {\em {Minimal Composite Dynamics
  versus Axion Origin of the Diphoton excess},}
\href{http://arxiv.org/abs/1512.05334}{{arXiv:1512.05334 [hep-ph]}}.

\bibitem{Pilaftsis:2015ycr}
A.~Pilaftsis, {\em {Diphoton Signatures from Heavy Axion Decays at the CERN
  Large Hadron Collider},}
  \href{http://dx.doi.org/10.1103/PhysRevD.93.015017}{Phys. Rev. {\bfseries
  D93} no.~1, (2016) 015017},
\href{http://arxiv.org/abs/1512.04931}{{arXiv:1512.04931 [hep-ph]}}.

\bibitem{Mambrini:2015wyu}
Y.~Mambrini, G.~Arcadi, and A.~Djouadi, {\em {The LHC diphoton resonance and
  dark matter},}
\href{http://arxiv.org/abs/1512.04913}{{arXiv:1512.04913 [hep-ph]}}.

\bibitem{Tetradis:2003qa}
N.~Tetradis, {\em {The Quark meson model and the phase diagram of two flavor
  QCD},} \href{http://dx.doi.org/10.1016/S0375-9474(03)01624-5}{Nucl. Phys.
  {\bfseries A726} (2003) 93--119},
\href{http://arxiv.org/abs/hep-th/0303244}{{arXiv:hep-th/0303244 [hep-th]}}.

\bibitem{Drees:1994zx}
M.~Drees, R.~M. Godbole, M.~Nowakowski, and S.~D. Rindani, {\em {$\gamma
  \gamma$ processes at high energy $pp$ colliders},}
  \href{http://dx.doi.org/10.1103/PhysRevD.50.2335}{Phys. Rev. {\bfseries D50}
  (1994) 2335--2338},
\href{http://arxiv.org/abs/hep-ph/9403368}{{arXiv:hep-ph/9403368 [hep-ph]}}.

\bibitem{Martin:2004dh}
A.~D. Martin, R.~G. Roberts, W.~J. Stirling, and R.~S. Thorne, {\em {Parton
  distributions incorporating QED contributions},}
  \href{http://dx.doi.org/10.1140/epjc/s2004-02088-7}{Eur. Phys. J. {\bfseries
  C39} (2005) 155--161},
\href{http://arxiv.org/abs/hep-ph/0411040}{{arXiv:hep-ph/0411040 [hep-ph]}}.

\bibitem{Berger:1983yi}
E.~L. Berger, E.~Braaten, and R.~D. Field, {\em {Large-$p_{T}$ production of
  single and double photons in proton-proton and pion-proton collisions},}
\href{http://dx.doi.org/10.1016/0550-3213(84)90084-1}{Nucl. Phys. {\bfseries
  B239} (1984) 52}.

\bibitem{Glover:1988fe}
E.~W.~N. Glover and J.~J. van~der Bij, {\em {Vector boson pair production via
  gluon fusion},}
\href{http://dx.doi.org/10.1016/0370-2693(89)91099-X}{Phys. Lett. {\bfseries
  B219} (1989) 488}.

\bibitem{Hahn:1998yk}
T.~Hahn and M.~Perez-Victoria, {\em {Automatized one loop calculations in
  four-dimensions and D-dimensions},}
  \href{http://dx.doi.org/10.1016/S0010-4655(98)00173-8}{Comput. Phys. Commun.
  {\bfseries 118} (1999) 153--165},
\href{http://arxiv.org/abs/hep-ph/9807565}{{arXiv:hep-ph/9807565 [hep-ph]}}.

\bibitem{vanOldenborgh:1989wn}
G.~J. van Oldenborgh and J.~A.~M. Vermaseren, {\em {New Algorithms for One Loop
  Integrals},}
\href{http://dx.doi.org/10.1007/BF01621031}{Z. Phys. {\bfseries C46} (1990)
  425--438}.

\bibitem{Lebiedowicz:2016ioh}
P.~Lebiedowicz, O.~Nachtmann, and A.~Szczurek, {\em {Central exclusive
  diffractive production of the $\pi^{+}\pi^{-}$ continuum, scalar and tensor
  resonances in $pp$ and $p \bar{p}$ scattering within the tensor Pomeron
  approach},} \href{http://dx.doi.org/10.1103/PhysRevD.93.054015}{Phys. Rev.
  {\bfseries D93} no.~5, (2016) 054015},
\href{http://arxiv.org/abs/1601.04537}{{arXiv:1601.04537 [hep-ph]}}.

\bibitem{Martin:2009iq}
A.~D. Martin, W.~J. Stirling, R.~S. Thorne, and G.~Watt, {\em {Parton
  distributions for the LHC},}
  \href{http://dx.doi.org/10.1140/epjc/s10052-009-1072-5}{Eur. Phys. J.
  {\bfseries C63} (2009) 189--285},
\href{http://arxiv.org/abs/0901.0002}{{arXiv:0901.0002 [hep-ph]}}.

\bibitem{Abazov:2013pua}
V.~M. Abazov {\em et~al.}, (D0 Collaboration), {\em {Measurement of the
  differential cross sections for isolated direct photon pair production in $p
  \bar p$ collisions at $\sqrt{s} = 1.96$ TeV},}
  \href{http://dx.doi.org/10.1016/j.physletb.2013.06.036}{Phys. Lett.
  {\bfseries B725} (2013) 6--14},
\href{http://arxiv.org/abs/1301.4536}{{arXiv:1301.4536 [hep-ex]}}.

\bibitem{Aad:2012tba}
G.~Aad {\em et~al.}, (ATLAS Collaboration), {\em {Measurement of
  isolated-photon pair production in $pp$ collisions at $\sqrt{s}=7$ TeV with
  the ATLAS detector},} \href{http://dx.doi.org/10.1007/JHEP01(2013)086}{JHEP
  {\bfseries 01} (2013) 086},
\href{http://arxiv.org/abs/1211.1913}{{arXiv:1211.1913 [hep-ex]}}.

\bibitem{Chatrchyan:2014fsa}
S.~Chatrchyan {\em et~al.}, (CMS Collaboration), {\em {Measurement of
  differential cross sections for the production of a pair of isolated photons
  in $pp$ collisions at $\sqrt{s}=7\,\text {TeV} $},}
  \href{http://dx.doi.org/10.1140/epjc/s10052-014-3129-3}{Eur. Phys. J.
  {\bfseries C74} no.~11, (2014) 3129},
\href{http://arxiv.org/abs/1405.7225}{{arXiv:1405.7225 [hep-ex]}}.

\bibitem{Aaltonen:2008dn}
T.~Aaltonen {\em et~al.}, (CDF Collaboration), {\em {Search for new particles
  decaying into dijets in proton-antiproton collisions at $\sqrt{s}$ = 1.96
  TeV},} \href{http://dx.doi.org/10.1103/PhysRevD.79.112002}{Phys. Rev.
  {\bfseries D79} (2009) 112002},
\href{http://arxiv.org/abs/0812.4036}{{arXiv:0812.4036 [hep-ex]}}.

\bibitem{Aad:2013tea}
G.~Aad {\em et~al.}, (ATLAS Collaboration), {\em {Measurement of dijet cross
  sections in $pp$ collisions at 7 TeV centre-of-mass energy using the ATLAS
  detector},} \href{http://dx.doi.org/10.1007/JHEP05(2014)059}{JHEP {\bfseries
  05} (2014) 059},
\href{http://arxiv.org/abs/1312.3524}{{arXiv:1312.3524 [hep-ex]}}.

\bibitem{Jia:2012kd}
J.~Jia, S.~Matsuzaki, and K.~Yamawaki, {\em {Walking technipions at the LHC},}
  \href{http://dx.doi.org/10.1103/PhysRevD.87.016006}{Phys. Rev. {\bfseries
  D87} no.~1, (2013) 016006},
\href{http://arxiv.org/abs/1207.0735}{{arXiv:1207.0735 [hep-ph]}}.

\bibitem{Harland-Lang:2016qjy}
L.~A. Harland-Lang, V.~A. Khoze, and M.~G. Ryskin, {\em {The production of a
  diphoton resonance via photon-photon fusion},}
  \href{http://dx.doi.org/10.1007/JHEP03(2016)182}{JHEP {\bfseries 03} (2016)
  182},
\href{http://arxiv.org/abs/1601.07187}{{arXiv:1601.07187 [hep-ph]}}.

\bibitem{Bern:2011ep}
Z.~Bern, G.~Diana, L.~J. Dixon, F.~Febres~Cordero, S.~Hoeche, D.~A. Kosower,
  H.~Ita, D.~Maitre, and K.~Ozeren, {\em {Four-Jet Production at the Large
  Hadron Collider at Next-to-Leading Order in QCD},}
  \href{http://dx.doi.org/10.1103/PhysRevLett.109.042001}{Phys. Rev. Lett.
  {\bfseries 109} (2012) 042001},
\href{http://arxiv.org/abs/1112.3940}{{arXiv:1112.3940 [hep-ph]}}.

\bibitem{Kutak:2016mik}
K.~Kutak, R.~Maciula, M.~Serino, A.~Szczurek, and A.~van Hameren, {\em
  {Four-jet production in single- and double-parton scattering within
  high-energy factorization},}
\href{http://arxiv.org/abs/1602.06814}{{arXiv:1602.06814 [hep-ph]}}.

\bibitem{Barrie:2016ntq}
N.~D. Barrie, A.~Kobakhidze, M.~Talia, and L.~Wu, {\em {750 GeV Composite Axion
  as the LHC Diphoton Resonance},}
  \href{http://dx.doi.org/10.1016/j.physletb.2016.02.010}{Phys. Lett.
  {\bfseries B755} (2016) 343--347},
\href{http://arxiv.org/abs/1602.00475}{{arXiv:1602.00475 [hep-ph]}}.

\bibitem{Kurachi:2014xla}
M.~Kurachi, S.~Matsuzaki, and K.~Yamawaki, {\em {Walking technipions in a
  holographic model},}
  \href{http://dx.doi.org/10.1103/PhysRevD.90.095013}{Phys. Rev. {\bfseries
  D90} no.~9, (2014) 095013},
\href{http://arxiv.org/abs/1403.0467}{{arXiv:1403.0467 [hep-ph]}}.

\bibitem{Fukano:2015hga}
H.~S. Fukano, M.~Kurachi, S.~Matsuzaki, K.~Terashi, and K.~Yamawaki, {\em {2
  TeV walking technirho at LHC?},}
  \href{http://dx.doi.org/10.1016/j.physletb.2015.09.023}{Phys. Lett.
  {\bfseries B750} (2015) 259--265},
\href{http://arxiv.org/abs/1506.03751}{{arXiv:1506.03751 [hep-ph]}}.

\bibitem{Aad:2015owa}
G.~Aad {\em et~al.}, (ATLAS), {\em {Search for high-mass diboson resonances
  with boson-tagged jets in proton-proton collisions at $\sqrt{s}=8$ TeV with
  the ATLAS detector},} \href{http://dx.doi.org/10.1007/JHEP12(2015)055}{JHEP
  {\bfseries 12} (2015) 055},
\href{http://arxiv.org/abs/1506.00962}{{arXiv:1506.00962 [hep-ex]}}.

\bibitem{Aad:2015ipg}
G.~Aad {\em et~al.}, (ATLAS), {\em {Combination of searches for $WW$, $WZ$, and
  $ZZ$ resonances in $pp$ collisions at $\sqrt{s} = 8$ TeV with the ATLAS
  detector},} \href{http://dx.doi.org/10.1016/j.physletb.2016.02.015}{Phys.
  Lett. {\bfseries B755} (2016) 285--305},
\href{http://arxiv.org/abs/1512.05099}{{arXiv:1512.05099 [hep-ex]}}.

\end{thebibliography}\endgroup

\end{document}